\documentclass{article} % For LaTeX2e
\usepackage{collas2022_conference,times}
\usepackage{amsmath,amsfonts,bm}

\def\eqref#1{equation~\ref{#1}}
\def\1{\bm{1}}

\DeclareMathAlphabet{\mathsfit}{\encodingdefault}{\sfdefault}{m}{sl}
\SetMathAlphabet{\mathsfit}{bold}{\encodingdefault}{\sfdefault}{bx}{n}

\usepackage{multicol}
\usepackage{wrapfig}
\usepackage{amssymb,amsfonts}
\usepackage{subcaption}
\usepackage{amsmath}
\usepackage{graphicx}
\usepackage{textcomp}
\usepackage{multirow}
\usepackage{csvsimple}
\usepackage{balance}
\usepackage{tikz}
\usepackage{todonotes}
\usepackage{tabularx}
\usepackage{xcolor}

\usepackage{enumitem}
\setitemize{noitemsep,topsep=0pt,parsep=2pt,partopsep=0pt,leftmargin=16pt}

\usepackage{hyperref}
\hypersetup{
    colorlinks=true,
    linkcolor=red,
    filecolor=magenta,      
    urlcolor=blue,
    citecolor=purple,
    pdftitle={Overleaf Example},
    pdfpagemode=FullScreen,
    }

\makeatletter
\renewcommand{\paragraph}{%
  \@startsection{paragraph}{4}%
  {\z@}{0ex \@plus 0.5ex \@minus .2ex}{-0.5em}%
  {\normalfont\normalsize\bfseries}%
}
\makeatother

\title{On the Limitations of Continual Learning for Malware Classification}

\author{Mohammad Saidur Rahman, Matthew Wright\\
ESL Global Cybersecurity Institute\\
Rochester Institute of Technology\\
{\small
\texttt{saidur.rahman@mail.rit.edu}, \texttt{matthew.wright@rit.edu}}\\
\And % Use And to have authors side by side
Scott E. Coull\\
Mandiant \\
{\small \texttt{scott.coull@mandiant.com}}
}

\collasfinalcopy % Uncomment for camera-ready version, but NOT for submission.

\begin{document}
%This document contains 

\maketitle

\begin{abstract}
Malicious software (malware) classification offers a unique challenge for 
continual learning (CL) regimes due to the volume of new samples received on a daily basis and the evolution of malware to exploit new vulnerabilities. On a typical day, antivirus vendors receive hundreds of thousands of unique pieces of software, both malicious and benign, and over the course of the lifetime of a malware classifier, more than a billion samples can easily accumulate. Given the scale of the problem, sequential training using continual learning techniques could provide substantial benefits in reducing training and storage overhead. To date, however, there has been no exploration of CL applied to malware classification tasks. In this paper, we study 11 CL techniques applied to three malware tasks covering common incremental learning scenarios, including task, class, and domain incremental learning (IL). Specifically, using two realistic, large-scale malware datasets, we evaluate the performance of the CL methods on both binary malware classification (Domain-IL) and multi-class malware family classification (Task-IL and Class-IL) tasks. To our surprise, continual learning methods significantly underperformed naive {\em Joint} replay of the training data in nearly all settings -- in some cases reducing accuracy by more than 70 percentage points. A simple approach of selectively replaying 20\% of the stored data achieves better performance, with 50\% of the training time compared to {\em Joint} replay. Finally, we discuss potential reasons for the unexpectedly poor performance of the CL techniques, with the hope that it spurs further research on developing techniques that are more effective in the malware classification domain.

\end{abstract}
\section{Introduction}
\label{intro}

Machine learning (ML) and deep learning (DL) models have become important tools in the defense of computers and networked systems. In particular, for addressing malicious software (\emph{malware}), ML-based approaches have been used extensively in both academic literature and real-world systems for malware detection and classification, 
including for Windows malware~\citep{tahan2012mal,dahl2013large, malwareguard}, PDF malware~\citep{maiorca2012pattern,laskov2011static}, malicious URLs~\citep{Lee2012WarningBirdDS,stringhini2013shady}, and Android malware~\citep{arp2014drebin,grosse2017adversarial,onwuzurike2019mamadroid}. These models are typically trained on previously observed samples and then deployed with the assumption that the model will generalize to new data. However, the adversarial nature of malware and continual evolution of benign software (\emph{goodware}) makes for an inherently non-stationary problem. 

To accommodate shifts in the data distribution over time, the model needs to be retrained regularly to maintain its effectiveness. Unfortunately, the speed with which new malware and goodware are produced results in large datasets that can be both costly to maintain and difficult to train on. For example, the AV-TEST institute registers more than 450,000 unique pieces of malware and ``potentially unwanted applications (PUA)" each day~\citep{av-test}, while VirusTotal, a crowdsourced antivirus scanning service, regularly receives more than 1 million unique pieces of software each day~\citep{virustotal}. Over the lifetime of a malware classification model, these daily feeds can result in datasets containing upwards of a billion unique training samples spanning multiple years. Given the realities of training these models, antivirus companies must decide whether to: (i) remove some older samples from the training set, at the risk of allowing attackers to revive older malware instead of writing new ones; (ii) train less frequently, at the cost of not adjusting to changes in the distribution; or (iii) expend tremendous effort to frequently retrain over all the data. 

Continual learning (CL) offers an appealing alternative to these options by enabling incremental incorporation of new information and adaptation to data distribution shifts without maintaining large datasets or incurring significant training overhead. 
In this work, we investigate the extent to which malware classification models suffer from catastrophic forgetting~\citep{mccloskey1989catastrophic, ratcliff1990connectionist, french1999catastrophic}, and whether we can address this using approaches from continual learning research. While combating the problem of catastrophic forgetting has been extensively studied, explored, and applied in the context of computer vision~\citep{gr,hsu2018re,rtf,bir,ji2007coordinated,farquhar2018towards} using the MNIST, CIFAR10 and CIFAR100, and ImageNet datasets, its role in and applicability to malware classification tasks remains unknown.

Using two large-scale malware datasets -- Drebin~\citep{arp2014drebin} and EMBER~\citep{ember} -- we examine the problems of binary malware classification and multi-class classification in the context of three CL scenarios: \emph{domain incremental learning (Domain-IL)}, \emph{class incremental learning (Class-IL)}, and \emph{task incremental learning (Task-IL)}. For Domain-IL, we focus on the binary classification task of labeling software as malicious or benign, and consider the problem of incorporating shifts in the data distribution over time without losing the ability to classify older samples. For Class-IL and Task-IL, we examine the task of malware \emph{family} classification, where a piece of malware is categorized into a well-defined family based on its code base, capabilities, and overall structure. In Class-IL, we incrementally add newly-discovered families at each iteration to mirror the ever-expanding universe of malware families found in the wild. The Task-IL formulation also incrementally adds new families to be classified, but constrains the classification task. All three scenarios represent problems faced in the anti-malware industry.

For each of these settings, we study 11 proposed CL approaches in our experiments, representing three major categories: regularization, replay, and replay with exemplars.
%: elastic weight consolidation (EWC)~\citep{ewc}, online EWC~\citep{onlineewc}, synaptic intelligence (SI)~\citep{si}, learning without forgetting (LwF)~\citep{lwf}, generative replay (GR)~\citep{gr} and a variant of GR paired with distillation (GR+Distill), replay through feedback (RtF)~\citep{lwf}, brain-inspired replay (BI-R)~\citep{bir}, experience replay (ER)~\citep{er}, averaged gradient episodic memory (A-GEM)~\citep{agem}, and incremental classifier and representation learning (iCaRL)~\citep{icarl}.} 
We investigate their ability to reduce catastrophic forgetting in our two malware datasets compared with the baselines of (i) using no CL techniques and (ii) full \emph{Joint replay}, which retrains on the full available dataset at each iteration. 
%In addition to these set of experiments, we have also investigated the extent to which Class-IL can be scaled to a higher number of classes using GR enhanced with different brain-inspired components~\citep{bir}~(Section~\ref{replaygenerateddata}). 
Finally, we investigate how much stored data may be enough to be replayed while still achieving high accuracy with lower storage and retraining costs. Our contributions are as follows:
\begin{itemize}
    \item We are among the first to investigate continual learning in security problems, specifically in malware classification using deep learning models.
    
    %\item We studied Domain-IL scenario using a real-world dataset with data distribution shift over the span of twelve months.
    
    \item We applied 11 distinct CL techniques in three CL scenarios using two large-scale malware datasets. We find that several CL techniques work reasonably well on Task-IL.
    %GR and GR with distillation, RtF, BI-R, and ER offer the best performance for Task-IL on the Drebin dataset, while LwF, ER, and A-GEM offer the best performance for Task-IL on the EMBER dataset.}     
    %offers the closest to the upper bound performance in Task-IL and Domain-IL scenarios, whereas GR yields closest to the upper bound performance in Class-IL scenario as long as the dataset has a small number of classes.
    
    \item We empirically show that none of the CL techniques are effective in the Domain-IL setting.
    
    \item For Class-IL, 10 of the 11 methods performed poorly, with only iCaRL~\citep{icarl} performing marginally better against the {\em Joint} replay baseline.
    %, suggesting that iCaRL suffers to retain its effectiveness when the dataset complexity is high. 
    
    \item Instead of storing all prior data, we find that replaying 20-50\% data is enough to significantly outperform all CL techniques in Domain-IL while also reducing the training cost by 35-50\% compared with 100\% {\em Joint} replay.  
    
    \item We discuss the probable causes of the poor performance offered by CL techniques to help spur future research in this problem setting. 
\end{itemize}

\section{Related Work}
\label{related_work}

Over the years, numerous mechanisms have been proposed to overcome catastrophic forgetting,
%~\citep{si,ewc,onlineewc,lwf,memoryawaresynapse,onlinelaplace,learningwmemorizing,rtf,bir} 
which fall into three major families~\citep{parisi2019continual}: i) \emph{regularization} methods, ii) \emph{replay} methods, and iii) \emph{adaptive expansion} methods. 

Regularization-based techniques attempt to penalize changes to weights that are determined to be important to the previous tasks. This is done by introducing a new loss function known as \emph{regularization loss}. The regularization loss is added to the training loss to compute the total loss. This category includes Elastic Weight Consolidation (EWC)~\citep{ewc}, EWC Online~\citep{onlineewc}, Synaptic Intelligence (SI)~\citep{si}, Memory Aware Synapses~\citep{memoryawaresynapse}, Riemannian Walk (RWALK)~\citep{riemannianwalk}, Online Laplace Approximator~\citep{onlinelaplace}, Hard Attention to the Task~\citep{hardattentiontask}, and Learning without Memorizing~\citep{learningwmemorizing}. %Among these methods, EWC, EWC Online, and SI are the most studied, and we thus adapt them for our experiments.

Replay methods are designed to complement the training data for each new task with older data that are representative of the tasks seen so far~\citep{chaudhry2019tiny,er}. Distillation loss, for instance, is often used in replay-based techniques~\citep{lwf,icarl,castro2018end}. Learning without Forgetting (LwF)~\citep{lwf} replays \emph{pseudo-data}, which is generated by first running the older model on new data to generate the softmax outputs, and then using those outputs as soft labels for training the model. This technique is similar to model distillation. %For each task, LwF trains with both the hard labels, to adapt to the new data distribution, and the soft labels, to retain aspects of the old model.

In an alternative form of replay called \emph{generative replay (GR)}~\citep{bir,rtf,gr}, a second model is trained to capture the distribution of data from the previous tasks and generate new samples from the distribution to be replayed. In the current task $T$, both the main model $\mathcal{M}_T$ and a generative model $\mathcal{G}_T$ is trained using a mix of task $T$ data and data generated by the prior generator model $\mathcal{G}_{T-1}$. GR is a powerful technique when access to older data is not available or restricted. Replay through Feedback (RtF)~\citep{rtf} reduces the training cost for dual-memory-based GR by coupling the generative model into the main model. RtF enhances the capability of the main model to generate samples by adding feedback connections and a layer of latent variables $z$, which are responsible for reconstructing inputs. Brain-Inspired Replay (BI-R)~\citep{bir} is an improvement over RtF and proposes three new components: i) replacement of the standard normal prior with Gaussian mixture, ii) internal context~\citep{xdg}, and iii) internal replay.

Another variant of replay, which intelligently picks the samples (i.e., called exemplars) to complement the training data of the current task, includes Experience Replay~\citep{er}, Gradient Episodic Memory (GEM)~\citet{gem}, Averaged Gradient Episodic Memory (A-GEM)~\citet{agem}, and Incremental Classifier and Representation Learning (iCaRL)~\citet{icarl}. ER leverages a reinforcement learning based learner to learn new and old experiences. In particular, ER balances stability and plasticity using on-policy for plasticity and off-policy for stability. iCaRL sets a memory budget beforehand, and the memory budget is equally divided into the previously learned classes. It picks and stores only exemplars that are close to the feature mean of each class. Loss of the current classes is minimized along with the distillation loss between targets obtained from the predictions of the previous model and the predictions of the current model on the previously learned classes. A-GEM also follows iCaRL to replay selective samples during training from the learned tasks to be replayed in the current task. When the original data is available, however, prior work has recommended \emph{exact replay} instead~\citep{icarl,nguyen2017variational,kemker2017fearnet}.

%\newtext{From the replay based techniques, we adapt ER, LwF, GR, a variant of GR in which the generated samples are labeled with \emph{soft targets} (i.e., output probabilities) instead of \emph{hard targets} (i.e., actual output labels) called as GR with Distillation (GR + Distill), RtF, and BI-R. In addition, we adapt A-GEM and iCaRL as the representative of exemplars-based methods.} 

%In an another set of experiments, where we assume that we have the older data to be replayed, we use exact replay.

Adaptive expansion methods grow the capacity of the neural network as new tasks are observed. Several techniques have been proposed, including Progressive Neural Networks~\citep{rusu2016progressive}, Dynamically Expandable Networks~\citep{yoon2017lifelong}, Adaptation by Distillation~\citep{hou2018lifelong}, and Dynamic Generative Memory~\citep{ostapenko2019learning}. These methods require incremental increases in memory as the new tasks are observed, and thus face scalability problems. Given the scale of malware classification problems, we leave the investigation of these techniques to future work, and instead focus on methods that do not require us to increase model capacity with dataset or problem size. 

Finally, we note that no prior work has explored continual learning to the malware domain. However, \citet{amalapuram2021handling} has proposed a continual learning-based network intrusion detection system (IDS) to mitigate catastrophic forgetting in a class-incremental scenario leveraging partial replay-based approaches. While they primarily focus on the role of class imbalance, some of the results of their study mirror our own findings, namely that the number and selection of samples used during replay are key to the success of continual learning in the cybersecurity space. We provide additional details and hypotheses for the underlying cause of this behavior given the unique nature of the malware classification problem in Section \ref{discussion}.

Additionally, there is some prior work on \emph{online learning} (OL) applied to malware classification~\citep{droidol, huynh2017new, narayanan2017context, droidevolver}. Online learning considers the problem of incorporating new samples into the model as they are observed, and notably does not directly address the problem of catastrophic forgetting. Furthermore, previous work has shown that producing high-quality labels for malware can be a difficult task~\citep{kantchelian2015better}, and often requires weeks of time for vendors to come to consensus on newly-discovered variants~\citep{zhu2020measuring}, making immediate incorporation of observed samples risky. Recently, researchers have explored malware detection systems leveraging transfer learning in an attempt to address the challenge of adapting to ever-evolving malware samples~\citep{hsiao2019malware, tang2020convprotonet, ale2020few, rong2021umvd, chai2022dynamic}. Transfer learning, however, is not focused on retaining knowledge of the prior tasks when being applied to the new domain. In malware, this distinction is significant, since an inability to detect previous malware variants would reintroduce vulnerabilities without the user's knowledge. We thus focus in this paper only on CL techniques, as they explicitly address the issue of catastrophic forgetting.
%Therefore, while OL and transfer-learning techniques are potentially applicable to the problems we study in this paper, we leave further exploration and comparison with CL methods to future work.

%choose to focus on practical concerns that have not yet been addressed by previous work.

% There is, however, some prior work in the space of {\em online learning} (OL) applied to malware classification~\citep{droidevolver, droidol}. OL considers ever increasing new samples over time which is closely similar to the domain incremental learning (Domain-IL). Even though OL approaches attempt to encounter future data in streaming or in a sequential order, the objective of OL is not to retain the knowledge learned so far to overcome catastrophic forgetting, rather just to update (retrain) the model with the new data so that the new model has knowledge of the new data. Another major limitation of OL is that it is not equipped to learn new classes or new tasks. Despite the limitations, OL can be a competitive technique to be applicable in the Domain-IL scenario coupled with some technique to retain the knowledge of the previous data distribution. 

\section{Preliminaries}
\label{preliminaries}

\subsection{Training Protocols}

\iffalse
\begin{itemize}
    \item $\theta_{s}$: parameters that are shared across all the tasks.
    \item $\theta_{0}$: parameters specific to the previous tasks.
    \item $\theta_{n}$: parameters specific to the new tasks.
\end{itemize}
\fi

Training in Continual learning (CL) involves considering three sets of parameters~\citep{lwf} -- i) $\theta_{s}$: parameters that are shared across all the tasks, ii) $\theta_{0}$: parameters specific to the previous tasks, and iii) $\theta_{n}$: parameters specific to the new task. Here, $\theta_{0}$ and $\theta_{n}$ contain the corresponding weights and the output layer of the older tasks and the new task, respectively. The weights of all the layers except the classification layer belong to $\theta_{s}$. 

Traditional deep learning (DL) training involves optimizing all three sets of parameters together, and is referred to as {\em Joint} training. This training method requires the availability -- and hence the storage -- of all the older data from previous tasks. The performance of this training mechanism is considered to be an upper bound, as data from all the previous tasks as well as the new tasks are used to train the model. Training this way, however, is slow and costly.

In CL training, for each new task $n$, $\theta_{s}$ and $\theta_{n}$ are optimized and updated while trying to keep $\theta_{0}$ fixed. CL training does not require the older data to be available, which makes this a difficult task.
%and thus the performance on the older task is usually bad.
Significant effort has been spent to boost the performance of models on the tasks in cases where there is no previous data available.
%or only an alternate version of the previous data available (i.e., generated data). 
Training is typically significantly faster than in {\em Joint} training, which may enable more frequent retraining of the model to keep up with changing data distributions or other needs without risking earlier concepts.

\subsection{Continual Learning Scenarios for Malware Classification}

We setup our continual learning problem in such a way that a ML model, $M$, learns a series of sequential tasks, $t_0, t_1,....,t_n$. During the training of $t_i$, only data from $t_i$ is available. 
% mkw - nothing was *formalized* below after this
Below, we describe three continual learning scenarios for malware classification~\citep{van2019three}: Domain Incremental Learning (Domain-IL), Class Incremental Learning (Class-IL), and Task Incremental Learning (Task-IL).

\paragraph{Domain-IL.} The most important problem in malware classification is binary classification of a test sample as either benign software (\emph{goodware}) or malware. Many new malware samples, unidentified applications, and goodware are created and released every day. New malware and goodware often behave differently from prior ones, resulting in concept drift over time. This makes it valuable to incorporate new samples into production systems sooner than later. Previously proposed malware classification techniques generally do not consider an evolving model trained continuously to incorporate this distribution shift without complete retraining~\citep{cade,transcend}. Further, in this adversarial setting, an attacker may leverage older malware specifically to evade classifiers that have forgotten about it. Thus, the distribution shift must be captured while avoiding catastrophic forgetting.

In this binary malware classification setting, we divide our dataset into monthly tasks, where each month captures the natural concept drift of both malware and goodware due to evolving malware capabilities and benign software releases. We seek to incorporate this new knowledge in each monthly incremental learning iteration while maintaining previous discriminative knowledge about earlier malware and goodware.

\emph{We note that, unlike most studies of Domain-IL that rely on artificial manipulations of image datasets, the malware dataset we use shows real-world shifts in the data distributions over time.} For example, we find that some malware families become more or less popular during the span of the dataset.
%This is similar to the Domain-IL concept, because we have a single classification task (discriminating between malware and goodware), where the underlying distribution shifts over time, and we want to ensure the model avoids catastrophically forgetting malware and goodware concepts months or even years later.

\paragraph{Class-IL.} The second type of malware task is multi-class family classification. A \emph{family} is a set of malware programs that have significant overlap in their code, such that they are considered by experts to be a group with common functionality. The famous Zeus banking trojan, for example, has evolved since 2006 to include 556 versions of software spread out among 35 different families with names like Citadel and Gameover~\citep{zeus}. 

It takes some time to label new malware samples and unknown applications with the help of experts and, in many cases, a consensus of scores from multiple anti-virus engines. Classes can be added when enough samples share enough similarity for the experts to decide that they are a new family~\citep{kantchelian2015better, zhu2020measuring}. This occurs fairly infrequently in practice, but it requires the model to be adaptable to incorporate this new knowledge. 

In this multi-class malware classification setting, we divide our dataset so that the model would learn new malware classes incrementally, extending its capabilities. We assume that the base model would start with a non-trivial number of classes, and then we would increment with new classes afterwards. Each new task is defined as adding new classes, but test time performance is measured on all classes that the model has been trained on so far.
%but at the test time the model does not know which classes are added at which task. 
%Class-IL gets harder with the increase of classes, as the model needs to retain the knowledge of all the older classes learned so far while adding the new task. 

In practice, an analysis pipeline is used to detect and triage malicious programs. The first step is binary classification of benign/malicious files. Once a file is determined to be malicious, it is useful to provide additional context to the security analyst about the malware family or its capabilities, which is encapsulated in Class IL. Since benign files are classified at the first stage in the pipeline, only malware is used in these settings.

\paragraph{Task-IL.}
To facilitate classification of malware into families, it can help to constrain the task based on information gained from other analysis methods, such as the broader category of the malware (e.g. adware, ransomware, etc.), behaviors of the malware~\citep{maliciousbehavior}, or the infection vector that the malware uses (e.g. phishing, downloader, etc.). Task-IL captures this notion of constrained tasks, where adding a new task may represent a new category or new set of behaviors. This is likely to be less frequent than simply adding a new family, as we have in Class-IL, but it represents a real problem in malware classification.
%Given an existing model already trained to examine features of malware based on millions of samples, it would be ideal to reuse that capability to address a new classification task. Expressed in this way, however, it maps more readily to multi-task learning or transfer learning than Task-IL.
%Task-IL as defined in the CL literature does not naturally map onto any real-world application in malware classification and analysis. In the CL literature, each new task involves training on new classes with new samples in a way that is more independent of the prior knowledge gained.
%Malware applications as described above would instead benefit from applying new labels to existing features, which is closer to a transfer learning setting. 
%Nevertheless, we examine the Task-IL setting for completeness, as it is commonly studied in CL~\citep{ewc, onlineewc, si}. Success in Task-IL may show at least some potential benefits for a CL technique in the malware domain.
%In some cases, it is helpful to know the category of the malware such as adware, ransomware, Trojan, and so on. Each category can contain multiple families of malware. This level of detail can assist an expert to investigate the sample in depth and to understand trends. Task-IL can provide the benefit to infer the families in each category of malware given the category. Each of the category can be each of the task and each task will contain their corresponding families. Ideally each category can have unequal number of families. 
Unlike in Class-IL, the task identity is given to the model at test time, making it a much easier problem. In malware, this could mean learning the task identity from a separate model, manual analysis, or field reports of the malware's behavior. As we do not have naturally defined tasks in our datasets, we divide our dataset into tasks that contain an equal number of independent and non-overlapping classes, as is common in the continual learning literature~\citep{ewc, onlineewc, si}.  
%This setting ideally represents the situation where we want to extend the capability of the model to learn new groups of malware families in which given the group identity (i.e., task identity) the model can infer which family the test sample belongs to. 
%Task-IL is the easiest form of incremental learning as task identity is given at the test time.

\subsection{Dataset}
We used two popular, large-scale datasets from the malware research community for our experiments: Drebin~\citep{arp2014drebin} for Android malware and EMBER~\citep{ember} for Windows {\em Portable Executable (PE)} malware. We use EMBER 2018, as this is the latest version and is designed to be more difficult for the classifier.

%msr --> we can use the table later if we have space
%\input{tables/drebin_data}
\paragraph{Drebin.}
The Drebin dataset consists of 5,560 Android malware samples.\footnote{\url{https://www.sec.cs.tu-bs.de/~danarp/drebin/download.html}} For our experiments, we use malware samples from the top 18 malware families totaling 4,525 malware samples. See~Appendix~\ref{drebindatastat} for more details. 

Drebin features are organized as sets of strings, such as permissions, API calls, and network addresses, and they are embedded in a joint vector space as Boolean expressions representing the presence or absence of the attribute. 
% I don't think this really works. First, MNIST is greyscale. And it's not nearly as sparse. And Drebin results are so different!
%This feature set may share some similarity to the features of MNIST dataset, where each pixel can be simplified to a Boolean expression -- on (white) and off (black), in which most of the features are set to zero. 
Drebin features are highly sparse, making it a relatively easy dataset. The total number of original features in these samples is 8,803. With binary features, we do not need to use standardization, though we do pre-process the features using \texttt{scikit-learn's} \texttt{VarianceThreshold}~\citep{pedregosa2011scikit} to omit features with very low variance ($<0.001$). This leaves a final set of 2,492 features. %The training, validation, and test split are 80:10:10 in our experiments.

We could only perform task-IL and class-IL experiments with this dataset. Even though the samples span over five years, the families in this dataset do not consistently contain enough samples for each time period.
%and the appeared families are not consistent over multiple periods. 

\paragraph{EMBER.}
The EMBER 2018 dataset contains features from one million PE files scanned mostly in 2018, with 50K samples from before 2018.\footnote{\url{https://github.com/elastic/ember}} There are 400K goodware samples, 400K malware samples, and 200K unknown samples. 

EMBER contain a variety of features, including general file information, header information, imported and exported functions, and section information. EMBER features also contain three groups of format-agnostic features: byte histogram, byte-entropy histogram, and string information. Some of these features with high cardinality (e.g., identified strings) are then processed using the \emph{hashing trick}~\citep{hashingtrick} with different bin sizes. Each of the groups of EMBER features have unique distribution characteristics, which makes this dataset complex.  It is also worth noting that the EMBER feature space encodes rich semantics for the underlying executable data that naturally constrain the feasible regions of that space. For more detail, refer to~\citet{ember}.

There are 2,381 features in total, and we use \texttt{scikit-learn's} \texttt{StandardScaler}~\citep{pedregosa2011scikit} to standardize the feature space. StandardScaler provides a \texttt{partial\_fit} method, which can be updated incrementally to standardize the dataset using each month representing a continuous flow of data.

For the Task-IL and Class-IL experiments, we only use the malware samples from 2018, which belong to 2,900 families. As we found that the majority of the families contain only a few samples, we filtered out the families containing fewer than 400 samples. This left us with 106 families, from which we select the top 100 malware families containing 337,035 samples for our experiments.

For the study of Domain-IL, we take both goodware and malware samples from 2018, removing the unknown samples (see~Appendix~\ref{emberdatastat}). This subset of the data spans 12 months, January to December. We focus on binary classification for this set of experiments.

\section{Experimental Settings}
\label{sec:approach}

\subsection{Model Selection and Training}
% 0.99642 ROCAUC LightGBM
% 99512 

%The features of both of the datasets in this study are manually crafted and extracted from raw bytes of goodware and malware samples. Developing a classifier with extracted features as input is relatively easier than with raw bytes as input. We use a multi-layer perceptron (MLP) for this study. 

%As EMBER is a relatively harder dataset with more samples than Drebin, we use an MLP model that provides us nearly to the state-of-the-art performance with EMBER dataset. 

For the purposes of our experiments, we standardize our model as a multi-layer perceptron (MLP). Our MLP model has four fully-connected (FC) layers with $[1024, 512, 256, 128]$ hidden units, using dropout ($rate = 0.5$) and batch normalization in each layer. We use SGD as the optimizer, with learning rate = $0.01$, momentum $=0.9$, and weight decay $=0.000001$. 
%In training, we use 30 epochs and 
%\texttt{Early Stopping} 
%early stopping with patience $=5$. 
To reach these values, we performed selective hyperparameter tuning, covering the number of layers, the number of hidden units, activation functions, optimization function, and learning rate.

This model attains an AUC score of $=0.99512$. As a baseline, we note that \citet{ember} built a LightGBM model for binary classification, and report a state-of-the-art AUC score of 0.99642 on EMBER 2018. 
%If we use \texttt{QuantileTransformer} instead of \texttt{StandardScaler} as the standardization method, our MLP model yields ROCAUC score = $0.99655$. We cannot use \texttt{QuantileTransformer} as it does not have \texttt{partial fit} technique. As our model's ROCAUC score with \texttt{StandardScaler} is very close to the ROCAUC score reported by \citet{ember} with lightGBM model, we finalize this model as for all the experiments with EMBER and Drebin. 

There are, however, a few data- and experiment-specific changes to the optimizer and learning rate that we finalized after conducting multiple sets of experiments in each setting. For experiments with the Drebin dataset, we observed that an Adam optimizer with learning rate $0.001$ provided us slightly better performance than SGD, so we switch to it for Task-IL and Class-IL experiments on Drebin. For Task-IL and Class-IL with the top 100 classes of EMBER, we observe that SGD with learning rate $0.001$ gave better accuracy than learning rate $0.01$. Meanwhile, for the Domain-IL experiments with EMBER, SGD with learning rate $0.01$ yielded the best performance. 
%Note that, in all of those experimental settings, we use the same MLP model architecture mentioned above.

For all scenarios and training protocols, the model is trained in a sequential manner, such that each of the $N$ tasks $t_1, t_2, t_3,...,t_N$ are independent, and their corresponding data distribution is also independent. The model only has access to the data of the current task. We use the standard multi-class cross-entropy loss for all the experiments except in Domain-IL. As Domain-IL is binary, we use binary cross entropy loss. We use mini-batch sizes of 32 and 256 for Drebin and EMBER, respectively.
%In all of the experimental settings, we train the model for 30 epochs.

\subsection{Implementation Details}

In Task-IL, we equally divide the number of classes into nine tasks for Drebin and and 20 tasks for EMBER data, where each task has two classes and five classes, respectively. In Class-IL with Drebin data, the model starts with 10 classes, and then we add two classes in each task, making five tasks in total. For EMBER, the model starts with 50 classes and then five classes are added in each task, making 11 tasks in total. In Class-IL, a task means an episode of learning new class(es). In Domain-IL, there are 12 tasks, one for each month's data of malware and goodware. 

The output layer of each of the scenarios is implemented differently. The output layers of Task-IL and Class-IL have a distinct output unit for each class to be learned. The major difference of these two scenarios, however, is in the {\em active} output units. In Task-IL, only the output units of the classes in the current task are active. In Class-IL, however, all the output units of all the classes seen so far are active. In Domain-IL, all the output units -- both of them in our binary classification task -- are active, as only the data distribution is changing. The \texttt{softmax} function only considers these {\em active} units while assigning probabilities. We performed each set of experiments 10 times using different random parameter initialization, and a random buffer population strategy to choose samples to be replayed from the stored data.

We used \texttt{PyTorch}~\citep{pytorch} and ran our models on a \texttt{CentOS-7} machine with an Intel Xeon processor with 40 CPU cores, 128GB RAM, and four GeForce RTX 2080Ti GPU cards, each with 12GB GPU memory. The code and processed dataset of this paper are available at \url{https://github.com/msrocean/continual-learning-malware/}.

\subsection{Baselines}

We have two baselines in this work: i) {\em None} and ii) {\em Joint}. In the {\em None} baseline, the model is trained sequentially without any CL techniques. The performance of this method can be interpreted as an informal lower bound on our results, though it is not a true lower bound in theory or practice. In the {\em Joint} baseline, the accumulated data of all the tasks observed so far are used to train the model. The performance of this mechanism can be interpreted as an informal upper bound, though it too is not a theoretical upper bound. {\em Joint} replay requires storage and training effort proportional to all the data of the tasks observed so far. It is expensive, but it ensures high performance across the entire dataset up to the current iteration. The effectiveness of a particular technique to overcome catastrophic forgetting can then be seen as its ability to move the accuracy from being near to the lower bound to near the upper bound.

\begin{table*}[t!]
\small
\def\arraystretch{1.1}
%\vskip 0cm
\vskip -1.0cm
\centering
\caption{\textbf{Summary of the Experiments.} The average accuracy (Mean) and minimum accuracy (Min) from all the tasks in each experiment. 
%Drebin dataset has 18 classes and Ember dataset has 100 classes for Task-IL and Class-IL experiments. 
%Ember Domain-IL considers binary classification. 
Results in \textbf{Bold} %, except {\em Joint} row, 
indicate accuracy values closer to {\em Joint} performance than {\em None}. {\bf EWC-O}: EWC Online, {\bf GR-D}: GR + Distill. Error range is omitted for the results with less than 1.0 standard deviation .}
%\textbf{Task-IL}: Task Incremental Learning, \textbf{Class-IL}: Class Incremental Learning, and \textbf{Domain-IL}: Domain Incremental Learning.
\vskip -0.25cm
\begin{tabular}{l|l|cc|cc|cc|cc|cc} 
\hline
\multirow[c]{3}{*}{\textbf{Approach}} & \multirow{3}{1em}{\textbf{Method}} & \multicolumn{4}{c|}{\textbf{Drebin}} & \multicolumn{6}{c}{\textbf{Ember}}\\ \cline{3-12} 
%\hline
%\rule{0pt}{2ex}
    &   & \multicolumn{2}{c|}{\textbf{Task-IL}} & \multicolumn{2}{c|}{\textbf{Class-IL}} & \multicolumn{2}{c|}{\textbf{Task-IL}} & \multicolumn{2}{c|}{\textbf{Class-IL}} & \multicolumn{2}{c}{\textbf{Domain-IL}} \\ \cline{3-12}
    
    & & Mean & Min & Mean & Min & Mean & Min & Mean & Min & Mean & Min  \\\hline    

%ember binary domain-IL
% 0.9262919948489688 0.8952653649904859
%0.9583727675454724 0.9408763323643411

\multirow{2}{2em}{Baselines}   &   None &    85.3   & 63.7$\pm$6  &    45.3   & 19.7 &    75.7   & 60$\pm$3.5 &   26.6  & 09.2 &    93.1 & 91.3 \\ 

 &   Joint &    \textbf{99.7}   & 99.3 &    \textbf{99.0}   & 97.5  &    \textbf{97.1} & 95$\pm$3   &   \textbf{87.7} &  85$\pm$2.5 &    \textbf{95.9} & 93.2 \\ \hline

\multirow{3}{2em}{Regul.}   &   EWC &    85.1   & 62$\pm$9 &    46.3   & 20$\pm$2 &    85.9   & 72$\pm$16 &    8.4   & 00.1 &    92.8 & 90.0 \\ 

 & EWC-O &    83.9   & 64$\pm$7 &    47.1   & 20$\pm$2 &    78.8   & 57$\pm$31 &    9.0   & 00.2 &    93.1 & 91.5 \\
 
 &   SI &    90.7   & 78$\pm$7 &    45.3   & 20.0 &   73.6   & 58$\pm$4 &    27.3   & 09.5 &    93.0   & 91.1 \\ \hline

\multirow[c]{5}{*}{Replay}   &   LwF &    \textbf{94.9}   & 88$\pm$4 &    27.8   & 5$\pm$1.5 &    \textbf{93.9}   & 91$\pm$9 &    11.9   & 00.7 &    93.2   & 91.7 \\ 

 &   GR &    \textbf{99.1}   & 98.1&    55.1   & 26.2 &    80.8   & 70$\pm$6 &    26.9   & 09.3 &    93.2   & 91.6 \\
 &   GR-D &    \textbf{99.3}   & 98.5 &    55.1   & 26.1 &    82.9   & 73$\pm$3 &    27.0   & 09.0 &    93.2   & 91.7 \\ 
 
 & RtF & {\bf 99.4} & 98.9 & 55.0 & 25.7 & 77.7 & 68$\pm$8.5 & 26.6 & 09.1 & 93.1 & 91.2 \\

 & BI-R & {\bf 95.9} & 88$\pm$6 & 58.7 & 30$\pm$2.5 & 86.9 & 81$\pm$4 & 26.7 & 9.0 & 93.4 & 91.6 \\ \hline  
 
\multirow{3}{4em}{Replay + Exemplars}
 & ER & {\bf 99.6} & 99.1 & 55.2 & 26.7 & {\bf 94.0} & 91$\pm$1 & 28.0 & 09.4 & 75.9 & 65$\pm$4.5 \\
    & A-GEM & 92.6 & 79$\pm$5.5 & 47.8 & 20$\pm$2 & {\bf 90.4} & 82$\pm$3 & 28.0 & 09.9 & 77.5 & 67.4 \\
    
    & iCaRL & - & - & {\bf 96.2} & 95$\pm$1 & - & - & {\bf 62.8} & 46$\pm$2.5 & - & - \\ 
     \hline  
%\rule{1pt}{3ex}
\hline
%\rule{0pt}{3ex}
\end{tabular}

\label{exp_summary}
\vskip -0.4cm   
\end{table*}

\subsection{Continual Learning Techniques Studied}
\label{adaptedtechniques}

In this work, we apply 11 widely studied CL techniques belonging to three major categories: regularization, replay, and replay with exemplars. 
%We apply three widely studied regularization-based continual learning techniques: i) Elastic Weight Consolidation (EWC)~\citep{ewc}, ii) EWC Online~\citep{onlineewc}, and Synaptic Intelligence (SI)~\citep{si}. Replay based technique in our experiments contain two most widely used methods: i) Learning without Forgetting (LwF)~\citep{lwf}, and ii) Generative Replay (GR)~\citep{gr}. We also use a variant of GR where ``soft targets'' are used instead of ``hard targets'' as the labels. 
We provide a brief overview of each of the studied techniques in this section. The details of each of these techniques are given in Appendix~\ref{adaptedtechniquesdetails}.  

\paragraph{Regularization.}
Elastic Weight Consolidation (EWC)~\citep{ewc} quantifies the importance of the weights in terms of their impact on the previous tasks' performance. 
% mkw - not defined nor common
%with Fisher Information matrix 
%and penalizes changes to parameters estimated to be important for previously learned tasks. 
As proposed, however, EWC is not scalable to a large number of tasks, as the regularization term grows with the number of tasks. EWC Online~\citep{onlineewc} is a modified version of EWC proposed to overcome this limitation. 
%EWC Online ensures that the computational cost of the regularization term does not grow with the increase of tasks by a more strict approximation of the Bayesian treatment that results in only a single quadratic penalty term on the parameters. 
Synaptic Intelligence (SI)~\citep{si} is similar to EWC Online, but it uses a different method to measure the importance of weights and a different regularization loss.
%in which only one quadratic term penalizes the changes to the parameters after training on the older task. To measure the importance of weights, SI replaces the Fisher Information matrix and introduces a new regularization loss called quadratic surrogate loss.

\paragraph{Replay.}
%Replay-based techniques attempt to \newtext{reuse data from the previous task, or} mimic replaying data from the previous tasks, while training on new tasks. 
%Learning without Forgetting (LwF)~\citep{lwf} replays \emph{pseudo-data}, which is generated by first running the older model on new data to generate the softmax outputs, and then using those outputs as soft labels for training the model. This technique is similar to model distillation. 
For each task, LwF trains with both the hard labels, to adapt to the new data distribution, and the soft labels, to retain aspects of the old model.
%which comes from the soft targets of data of the current tasks using the stored model of the previous task. 
%{\bf ER, RtF, BI-R}
Generative Replay (GR)~\citep{gr}, on the other hand, replays representative data of the previous tasks using a generative model.
%, meaning that real data from the past tasks do not need to be stored.
%The generated data works like an abstract representation of the information stored in the hippocampus. In GR, there are 
%In the current task $T$, we train both the main model $\mathcal{M}_T$ and a generative model $\mathcal{G}_T$ using a mix of task $T$ data and data generated by the prior generator model $\mathcal{G}_{T-1}$.
%to generates a representative of the data of the current task. 
%The GR~\citep{gr} framework is designed in such a way that choice of the generative model is not limited to a Generative Adversarial Network (GAN)~\citep{goodfellow2014generative} and can instead be a variational autoencoder (VAE)~\citep{kingma2013auto} or any other such type. 
For the generative model, $\mathcal{G}$, in our experiments, we use a symmetric VAE~\citep{kingma2013auto}, in which the base model architecture is used for both the encoder and the decoder~\citep{rtf, bir}. %The data to be replayed are sampled from the generative model, and then the selected samples are fed to the main model and then labeled based on the predicted class of the model. The samples to be replayed during task $T$ are generated by the version of the main model and the generator after training on task $T-1$. Hence, we need to store a copy of both $\mathcal{M}$ and $\mathcal{G}$ after each task. 
We use a variant of GR with distillation loss, as well. %In this approach, we label the generated samples using the full vector of softmax outputs from the main model, rather than a single hard label.
We also evaluate Replay through Feedback (RtF)~\citep{rtf}, which attempts to reduce the extensive cost of training the dual-memory-based GR technique by integrating the generative model into the main model. In addition, we study Brain-Inspired Replay (BI-R)~\citep{bir} which improves upon RtF.
%the ``soft target'' instead of ``hard target'' of the generated data using the main model.

\paragraph{Replay + Exemplars.}
Experience replay (ER)~\citep{er} trains the model to learn new experiences (i.e., new tasks) coupled with replayed experiences (i.e., old tasks). iCaRL~\citep{icarl} proposes to store $\mathcal{X}$ number of samples of the previously learned classes based on a memory budget. A-GEM~\citep{agem} contains an episodic memory which stores a subset of the observed examples from previously learned task and replayed along with the new sample during training.

%iCaRL stores images from earlier classes for replay, uses a distillation loss to preserve weights, and uses a nearest class mean classifier in feature space.

\iffalse
\begin{figure*}[!t]
%\vskip -1.0cm
\centering
\includegraphics[scale=0.35]{figures/Ember_CI_100_Task_IL_MLP.eps}
%\vskip 0.2cm    
\caption{\textit{\bf EMBER Task-IL}: Accuracy of different continual learning methods as the number of tasks learned grows}
\label{fig:ember_exp_task}
%\vspace{-0.4cm}
\end{figure*}
\fi

\if 0
LwF labels the current data using the model of the old task. This forms an input, output pair. The outputs in this case is not the actual label rather the predicted softmax probabilities of the current data using the old model. This input, output pairs are replayed while training the old model with the current data. Basically two things are happening here. i) we get the input, output pair of the current data using the old model, ii) we update the old model with the current data with actual input, output pairs and the input, output pairs we got from (i).  
\fi

\section{Evaluation}
\label{evaluation}

% of different CL techniques as the data distribution changes in each task month
\begin{figure}[!t]
%\vskip -1.0cm
\centering
\begin{minipage}[c]{0.32\linewidth}
\centering
\includegraphics[scale=0.28]{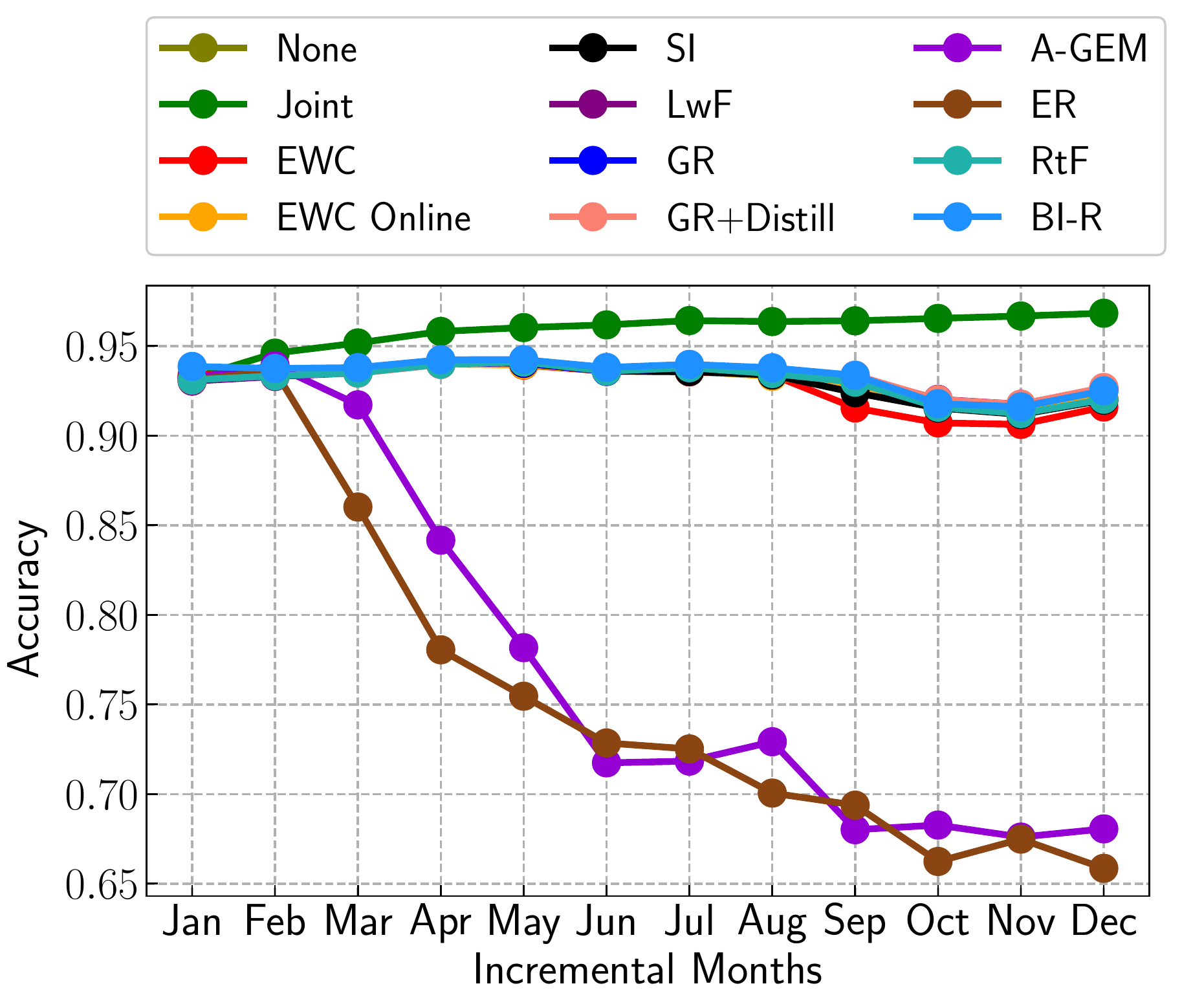}
\vskip -0.25cm %EMBER_Domain_Binary_MLP
\caption{\textit{\bf Domain-IL on EMBER}: Accuracy over time.}
\label{fig:domain_il_exps}
\end{minipage}%
\hfill
\begin{minipage}[c]{0.32\linewidth}
\includegraphics[scale=0.28]{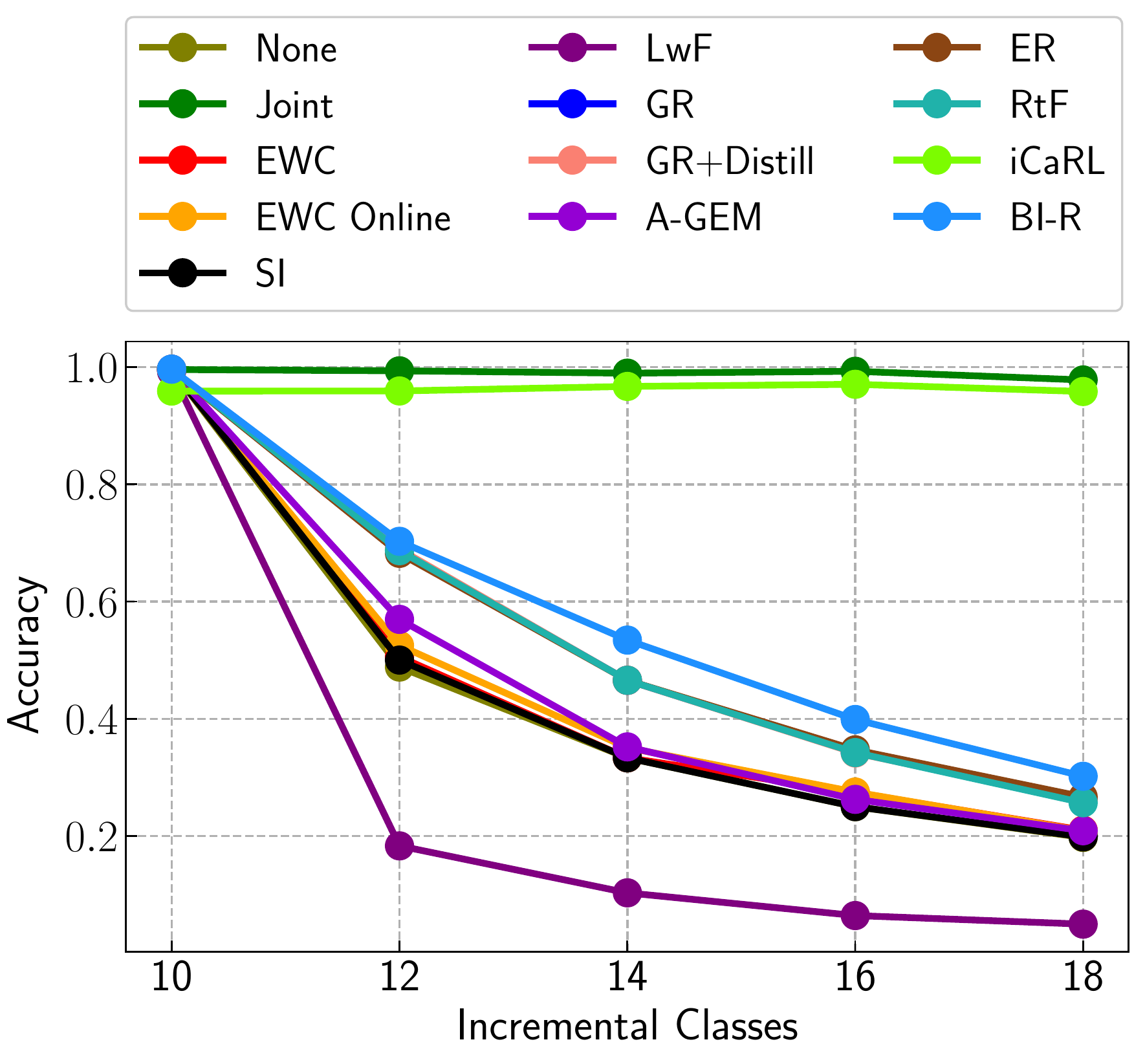}
\vskip -0.25cm
\caption{\textit{\bf Class-IL on Drebin}: Accuracy as the number of classes grows.}
\label{fig:drebin_exp_class}
\end{minipage}
\hfill
\begin{minipage}[c]{0.33\linewidth}
\centering
\includegraphics[scale=0.28]{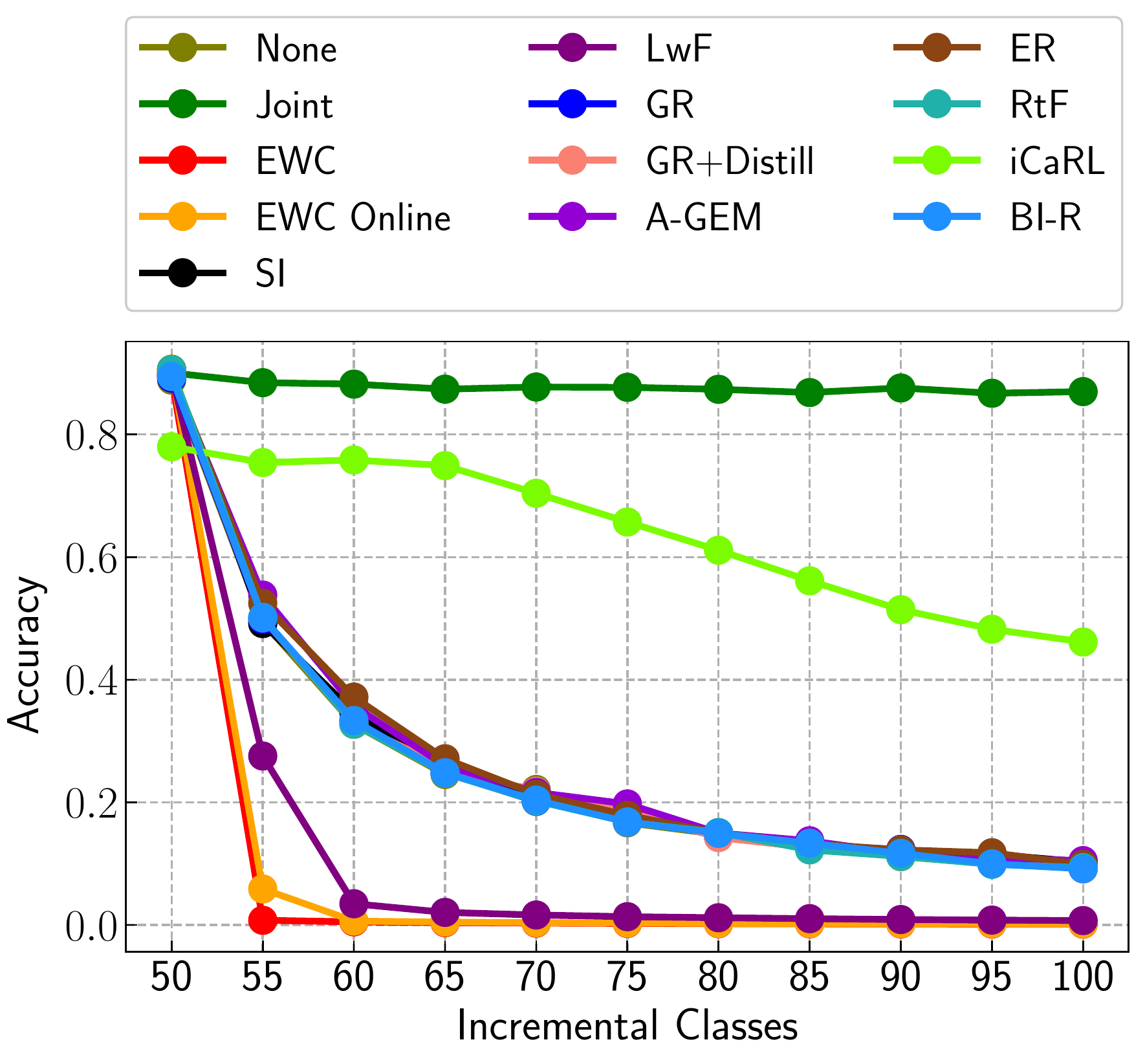}
\vskip -0.25cm
\caption{\textit{\bf Class-IL on EMBER}: Accuracy as the number of classes grows.}
\label{fig:ember_exp_class}
\end{minipage}%
%\vskip -0.3cm
\end{figure}

\subsection{CL Experiments}
In this section, we describe the results from our
main set of experiments,
%The first set of experiments (Section~\ref{CL-results}) 
investigating the three continual learning (CL) scenarios -- Domain-IL, Class-IL, and Task-IL -- using 11 continual learning techniques described in Section~\ref{adaptedtechniques}. %-- %EWC~\citep{ewc}, EWC Online~\citep{onlineewc}, SI~\citep{si}, LwF~\citep{lwf}, GR~\citep{gr}), and GR with distillation. 
%EWC, EWC Online, SI, LwF, GR, GR with distillation, LwF, BI-R, ER, A-GEM, and iCaRL. 
We compare the results with two baselines -- {\em None} and {\em Joint} -- and perform experiments using two malware datasets -- %Drebin~\citep{arp2014drebin} and EMBER~\citep{ember}. 
Drebin and EMBER.
We summarize our findings in Table~\ref{exp_summary}. The results of each of the experiments are represented in both {\bf Mean} and {\bf Min} metrics. Mean represents the mean accuracy of all the tasks in a single experiment, such as across 10, 12, 14, 16, and 18 classes tested for Class-IL with a given CL method on Drebin. Similarly, Min represents the minimum accuracy among all the tasks, which we highlight because the weakest performance shows the degree to which a technique may not be suitable for use.

%\subsection{Domain-IL}\label{domain_il}

\paragraph{Domain-IL.} In our experiments, we have 12 tasks representing the monthly data distribution shift of malware and goodware in EMBER from January to December 2018. Figure~\ref{fig:domain_il_exps} shows the results, which are summarized in the rightmost column of Table~\ref{exp_summary}. The mean accuracies of {\em None} and {\em Joint} baselines over the 12 tasks are 93.1\% and 95.9\%, respectively. We can see 
%from Figure~\ref{fig:domain_il_exps} 
that {Joint} performance trends upward with each incremental task. This may be expected, despite changes in the data distribution, as there is more training data in each additional task. We also see that none of the CL techniques are effective, with all of them performing closer to {\em None} than to {\em Joint}. Note that the data distribution does not change dramatically during the year -- even {\em None} reaches at least 91.3\% or better in all cases.
%in both regularization and replay categories contribute significantly to boost up the performance towards Joint level performance.

%\subsection{Class-IL}\label{class_il}
\paragraph{Class-IL.}
Figure~\ref{fig:drebin_exp_class} and Figure~\ref{fig:ember_exp_class} show the results of our experiments in this scenario for Drebin and EMBER, respectively. Since Class-IL is the most difficult CL scenario, it is not surprising that the mean accuracies for {\em None} and {\em Joint} are so far apart at 45.3\% and 99.0\%, respectively, on Drebin.
%The average accuracies of Drebin dataset across all five tasks on the {\em None} and {\em Joint} baselines are 45.60\% and 99.30\%. 
All the regularization-based techniques 
%(i.e., EWC, Online EWC, and SI) 
and LwF perform poorly and very close to {\em None}.
Among replay techniques, BI-R performs the best with 58.7\% mean accuracy. iCaRL outperforms all the other CL techniques with 96.2\% mean accuracy.
%As iCaRL utilizes the exemplars from the replay buffer from the previously learned classes, it significantly reduces catastrophic forgetting.}
%, which are the only two that provide better than {\em None} level performance. 
%With EMBER dataset, the average accuracies across all eleven tasks on the {\em None} and {\em Joint} baselines are 25.70\% and 85.10\%. The performances of all the CL techniques in both categories are much worse for EMBER dataset. However, SI stands out with a relatively higher accuracy than {\em None}. 
On EMBER, the performance of all methods is about the same or even worse than {\em None} except iCaRL, which yields 62.8\% mean accuracy. %As mentioned earlier, iCaRL replays the samples of the earlier classes stored in the memory buffer and the picked sampled are drawn based on the closeness to the mean feature factor of the classes which in turn reduces catastrophic forgetting significantly. This technique can be interpreted as an advanced version of partial joint replay.
%The complex and sparse feature space of EMBER might be one of the reasons for this level of low level performance in other techniques. 
Even though iCaRL outperforms other techniques by a substantial margin, it still is far from the {\em Joint} baseline, which is notable given the volume of executables scanned by malware detection models each day -- even a small increase in false positive or false negative rate can cause significant operational problems.

\begin{figure}[!t]
\vskip -.250cm
\begin{minipage}[c]{0.3\linewidth}
\centering
\includegraphics[width=\linewidth]{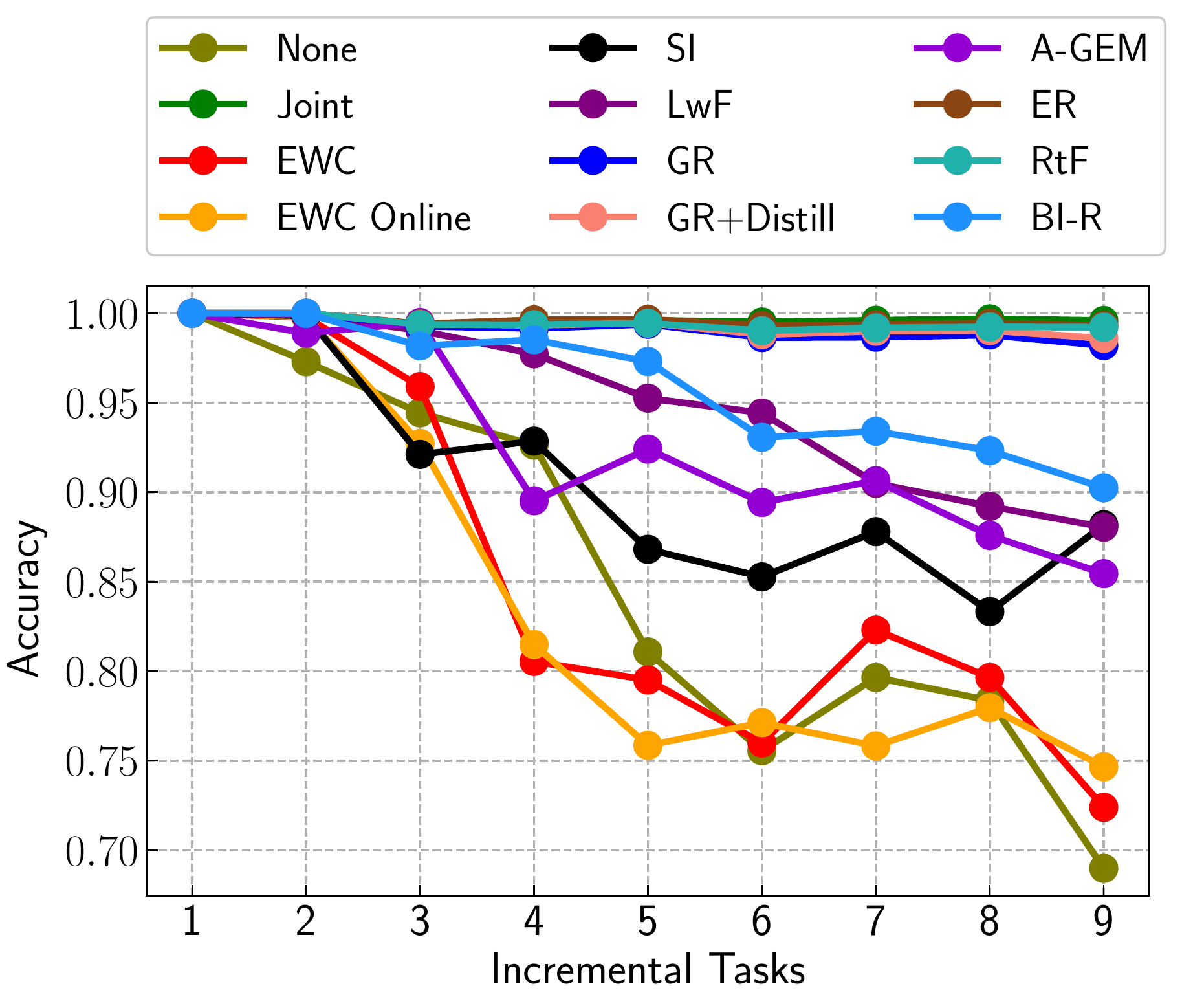}
\vskip -0.25cm
\caption{\textit{\bf Task-IL on Drebin}: Accuracy as number of tasks grows.}
\label{fig:drebin_exp_task}
\end{minipage}
\hfill
\begin{minipage}[c]{0.68\linewidth}
\centering
\includegraphics[width=\linewidth]{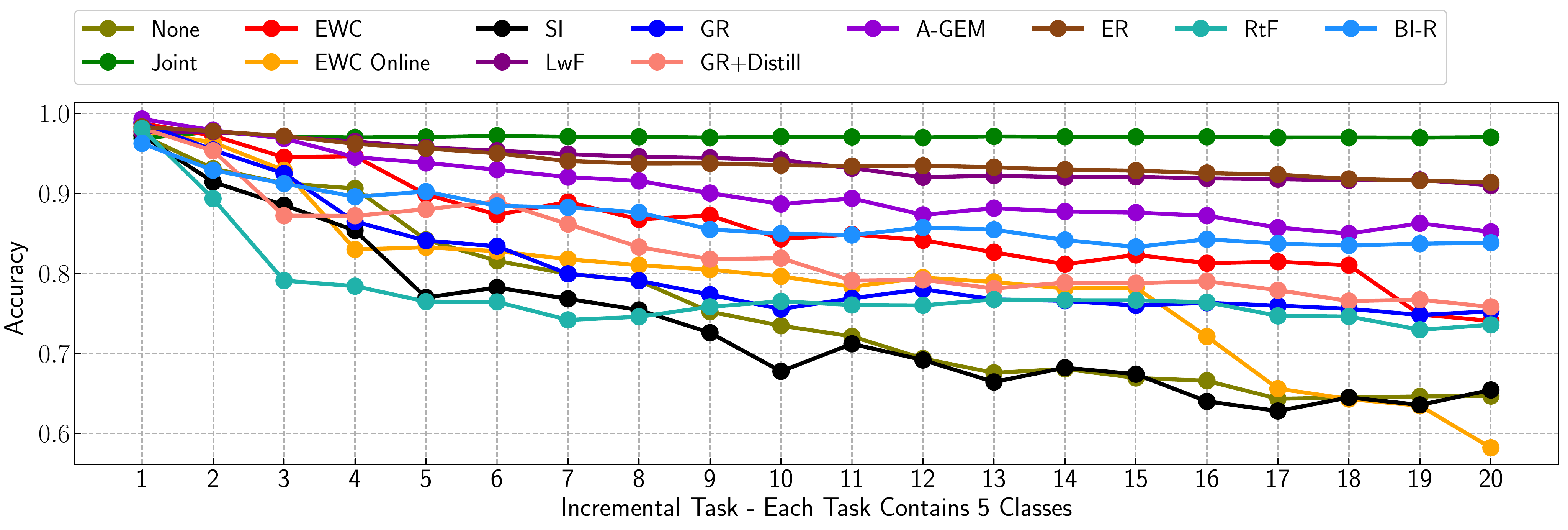}
\vskip -0.25cm
\caption{\textit{\bf Task-IL on EMBER}: Accuracy as the number of tasks grows.}
\label{fig:ember_exp_task}
\end{minipage}%
%\vskip -0.1cm
\end{figure}

%\subsection{Task-IL}\label{task_il}
\paragraph{Task-IL.}
Figures~\ref{fig:drebin_exp_task} and \ref{fig:ember_exp_task} show the results of the Task-IL experiments on Drebin and EMBER, respectively. The average accuracies on the Drebin dataset of {\em None} and {\em Joint} training of the nine tasks, where each task contains two classes, are 85.3\% and 99.7\%, respectively. These form the effective lower and upper bounds in this setting. Among the regularization techniques, only SI yields closer to {\em Joint} level performance with 90.7\% accuracy. The replay and replay-with-exemplars techniques are mostly effective, with GR, GR+Distill, RtF, and ER reaching over 99.0\% average accuracy.
%which are significantly closer to the {\em Joint} level performance. We can also observe that distillation in GR provides some advantage over GR. 

On EMBER, the average accuracy of {\em None} and {\em Joint} on the 20 tasks, where each task contains five classes, are 75.7\% and 97.1\%, respectively. None of the regularization-based techniques yield closer to {\em Joint} baseline performance. Among replay-based techniques, only LwF performs close to the {\em Joint} baseline with 93.9\% mean accuracy. Replay-with-exemplars-based techniques -- ER and A-GEM -- outperform other techniques with 94.0\% and 90.4\% average accuracy, respectively.
%EWC Online does reasonably well until task 15, but it fails badly afterwards. 
Considering both datasets, we can see that only ER performs reasonably well on both. GR, GR+Distill, RtF, BI-R, and SI all have difficulty with the larger and more complex EMBER dataset. In contrast, LwF does well on EMBER, but underwhelms on Drebin.
%while worked well for small number of tasks and simple dataset, break significantly to increasing tasks and dataset complexity.

\begin{figure}[!t]
\vskip -.250cm 
\begin{minipage}[c]{0.49\linewidth}
\centering
\includegraphics[scale=0.35]{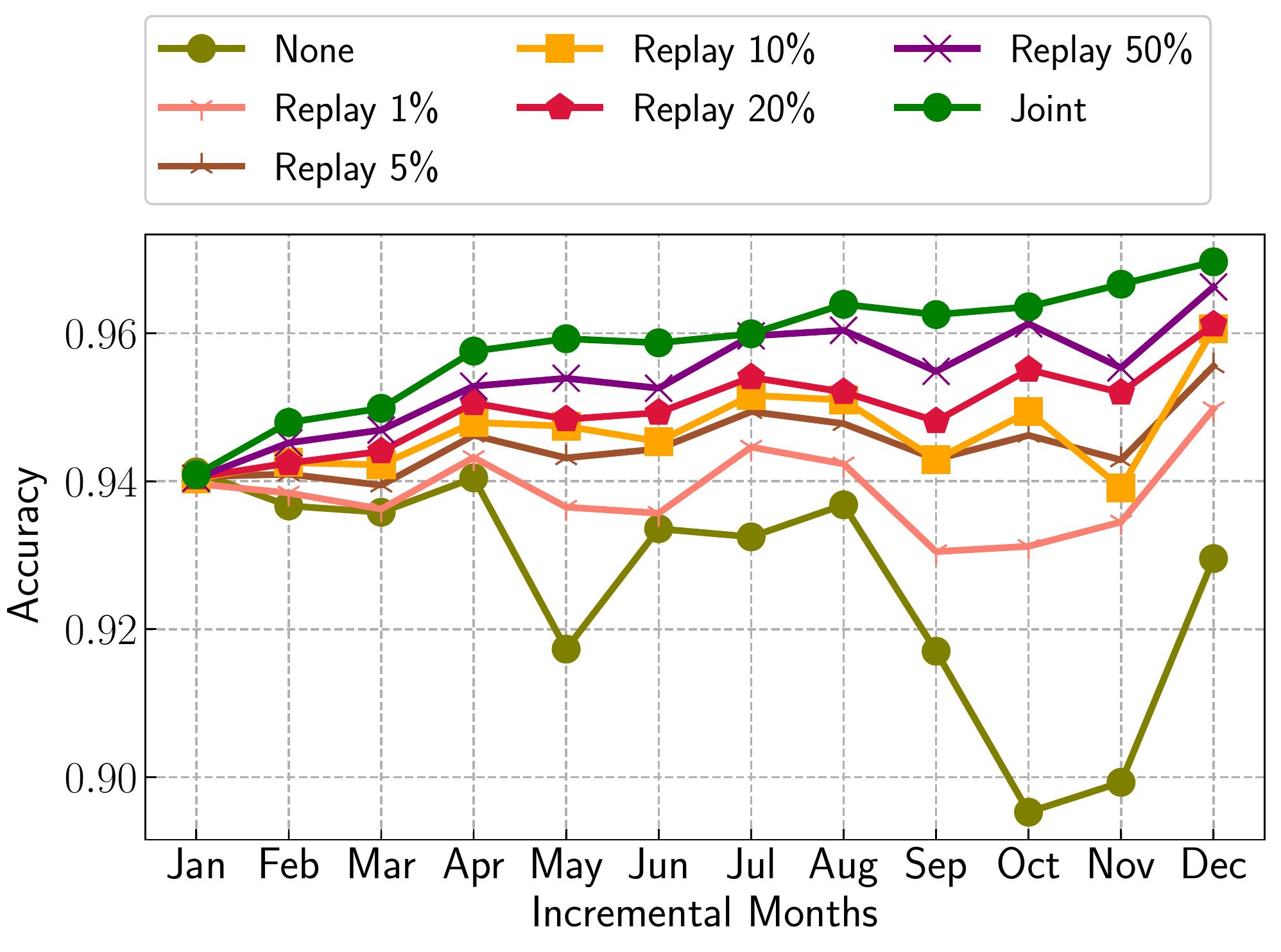}
\vskip -0.3cm
\caption{\textit{\bf Partial Joint Replay in EMBER}: Accuracy over time.}
\label{fig:domain_partial_il_exps}
\end{minipage}%
\hfill
\begin{minipage}[c]{0.49\linewidth}
\centering
\includegraphics[scale=0.35]{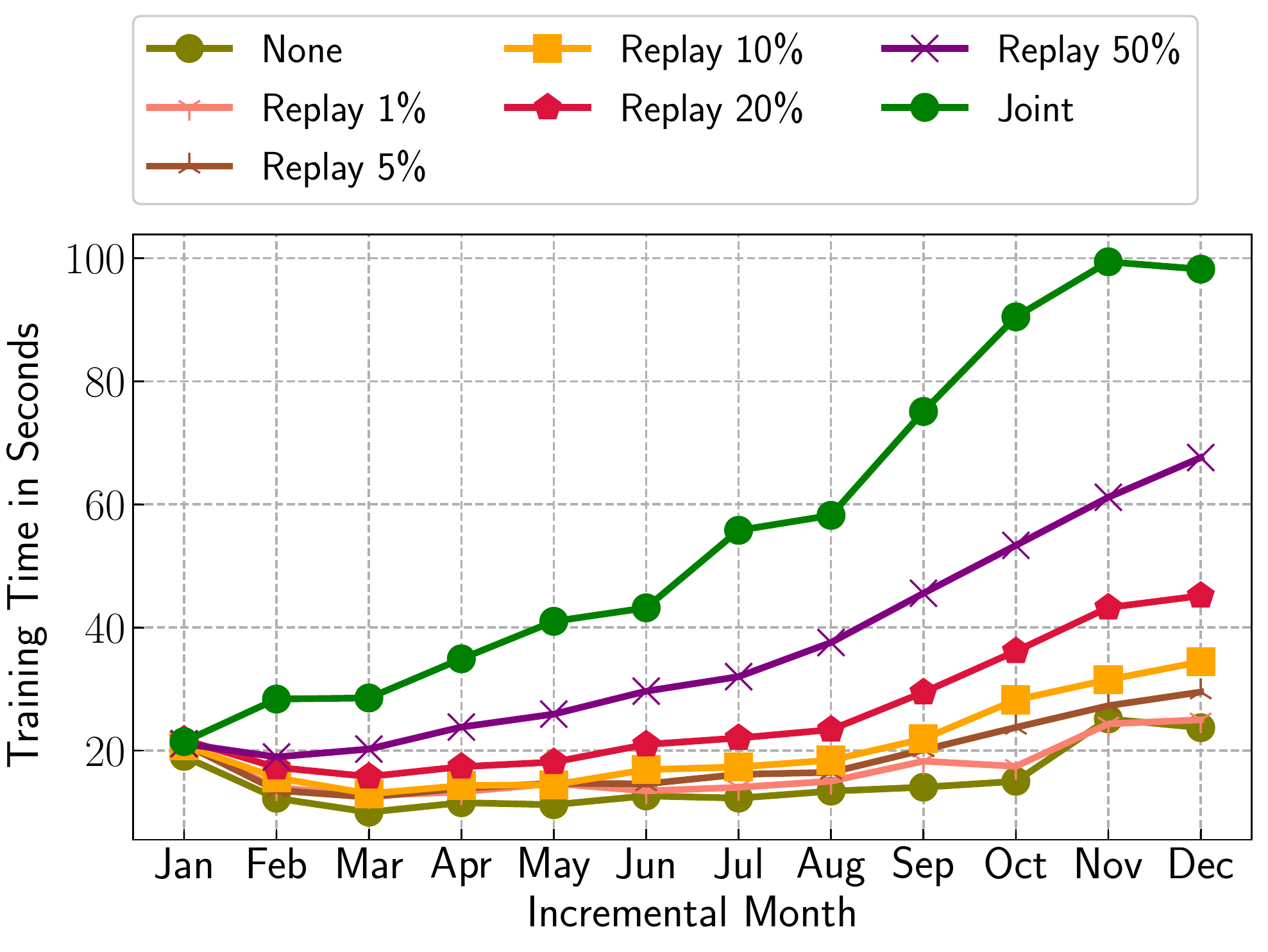}
\vskip -0.3cm
\caption{\textit{\bf Partial Joint Replay in EMBER}: Change in training time over time.}
\label{fig:domain_partial_il_training_time_exps}
\end{minipage}
\vskip -0.5cm
\end{figure}

\subsection{Partial Replay with Stored Data}
\label{replayrealdata}

In this section, we perform a second set of experiments using \emph{partial joint replay (PJR)}. This assumes that we have ample storage capacity and can maintain huge amounts of both historical and current data to be accessed at any time. Indeed, this may be the case for some companies doing malware detection. In this setting, CL techniques are not required, as the real data is available. Nevertheless, the huge volume of historical malware data makes it very expensive to train over all of it using full {\em Joint} replay. Thus, our question is how much of the historical data is needed to achieve high levels of accuracy using a strategy of sampled partial joint replay. In these experiments, we use early stopping with patience $= 5$, as replaying stored data causes the model to converge faster. 
%Hence using a fixed number of epochs to train the model in this setting will not provide a realistic measure of the training time.

We focus on Domain-IL, which is the CL setting most applicable to malware classification. We performed seven sets of experiments with varying fractions -- {\em None} (0\%), 1\%, 5\%, 10\%, 20\%, 50\%, and {\em Joint} (100\%) -- of the stored data to be replayed in the subsequent tasks. Figure~\ref{fig:domain_partial_il_exps} shows the results, while Figure~\ref{fig:domain_partial_il_training_time_exps} shows the corresponding training times.
%The expected upper limit for accuracy should be given by the {\em Joint} baseline which replays 100\% of the stored data. 
The mean accuracy of the {\em Joint} and {\em None} baselines is 96.4\% and 93.0\%, respectively. 
%The {\em None} baseline for this set of experiments is 93\%.

The average accuracy over all the tasks with 1\% of replayed data is 93.9\% -- nearly 1\% higher than {\em None} and easily outperforming all CL techniques evaluated above. It is perhaps surprising to see that only 1\% replayed data results in such a significant improvement in the average accuracy, but note that the total set of old training samples grows larger with each task, making it \emph{multiple times larger than the new data} for most tasks in our experiment. Even a small sample of 1\% thus becomes a significant fraction of all training data. When we increase the replayed data fraction to 5\%, 10\%, 20\%, and 50\%, average accuracy grows to 94.5\%, 94.7\%, 95.0\%, and 95.4\%, respectively. With 20\% replayed data, the accuracy is only 1.4\% lower than {\em Joint} replay. Reducing the replay data improves training efficiency significantly, as we can see in Figure~\ref{fig:domain_partial_il_training_time_exps}. With our dataset, the expected amount of training effort using 20\% of the replay data shrinks by 50\% compared to full {\em Joint} replay. Even with 50\% of the replay data, the expected training effort shrinks by 35\%.

\section{Discussion}
\label{discussion}

\begin{figure}[!t]
%\vskip -0.50cm
\centering
\begin{subfigure}[c]{0.16\linewidth}
\centering
\includegraphics[scale=0.11]{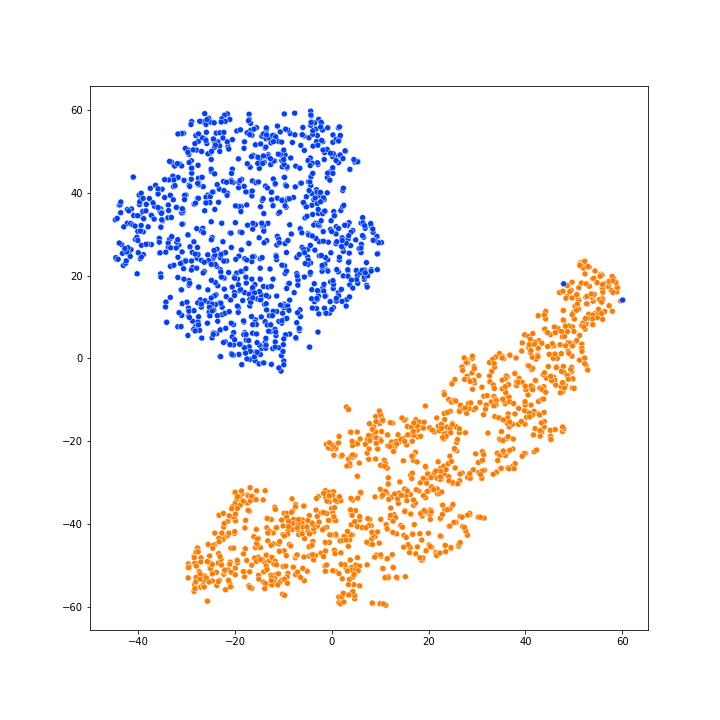}
%\vskip -0.5cm
\caption{MNIST 2-Class}
\label{fig:mnist_class_2_tSNE}
\end{subfigure}%
\hfill
\begin{subfigure}[c]{0.16\linewidth}
\centering
\includegraphics[scale=0.11]{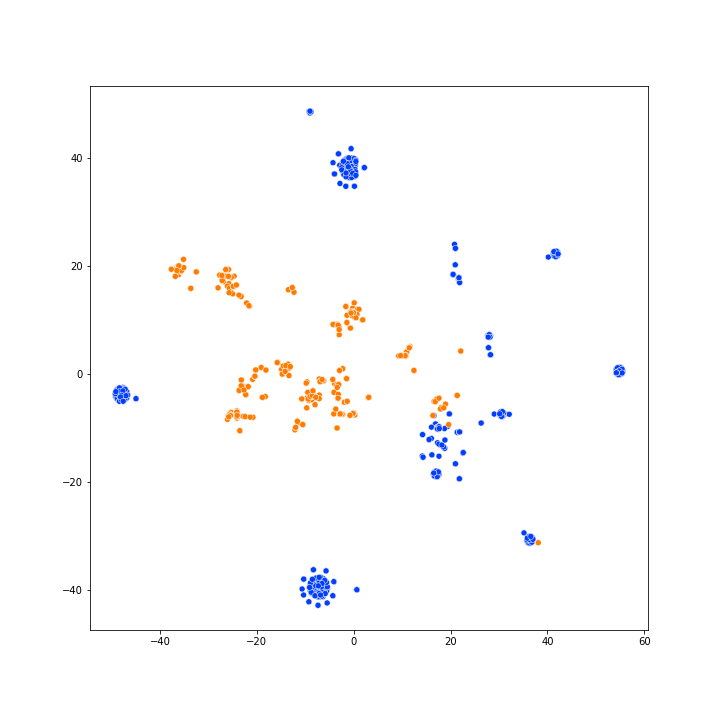}
%\vskip -0.5cm
\caption{Drebin 2-Class}
\label{fig:drebin_class_2_tSNE}
\end{subfigure}%
\hfill
\begin{subfigure}[c]{0.16\linewidth}
\centering
\includegraphics[scale=0.11]{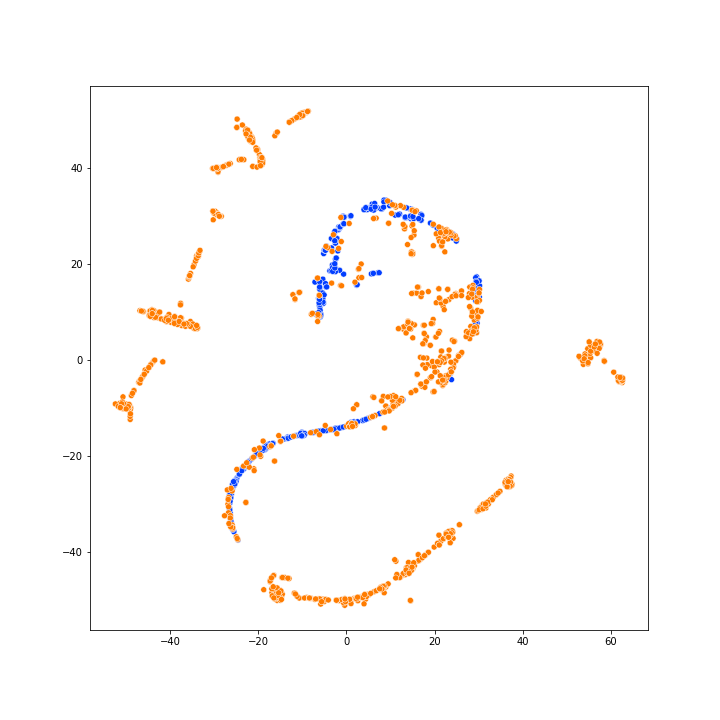}
%\vskip -0.5cm
\caption{EMBER 2-Class}
\label{fig:ember_ember_2_tSNE}
\end{subfigure}%
%\vfill
%\vskip -0.25cm
\hfill
\begin{subfigure}[c]{0.16\linewidth}
\centering
\includegraphics[scale=0.11]{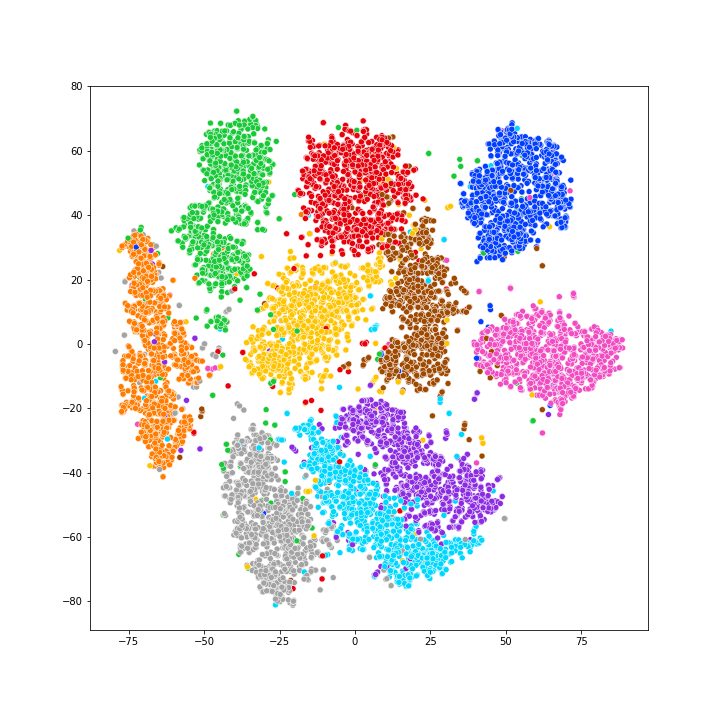}
%\vskip -0.5cm
\caption{MNIST 10-Class}
\label{fig:mnist_class_10_tSNE}
\end{subfigure}%
\hfill
\begin{subfigure}[c]{.16\linewidth}
\centering
\includegraphics[scale=0.11]{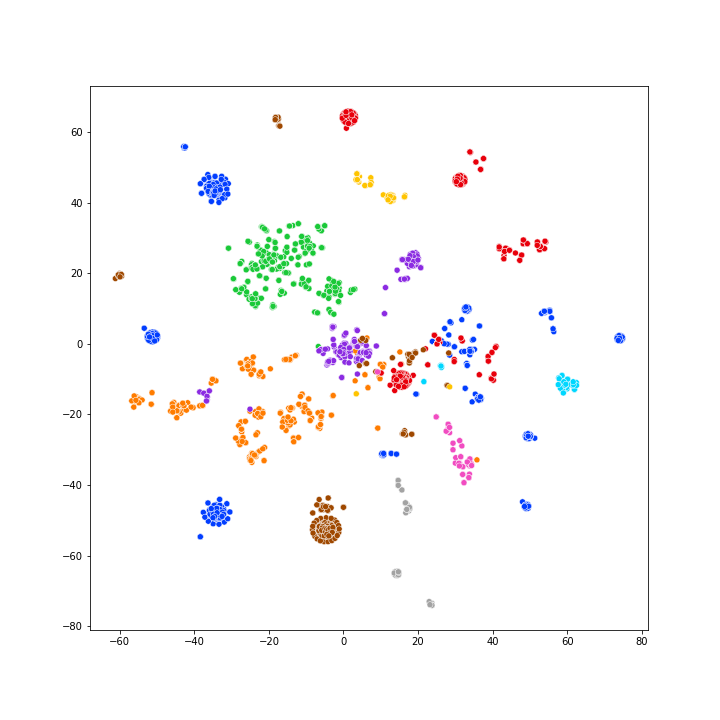}
%\vskip -0.5cm
\caption{Drebin 10-Class}
\label{fig:drebin_class_10_tSNE}
\end{subfigure}%
\hfill
\begin{subfigure}[c]{0.16\linewidth}
\centering
\includegraphics[scale=0.11]{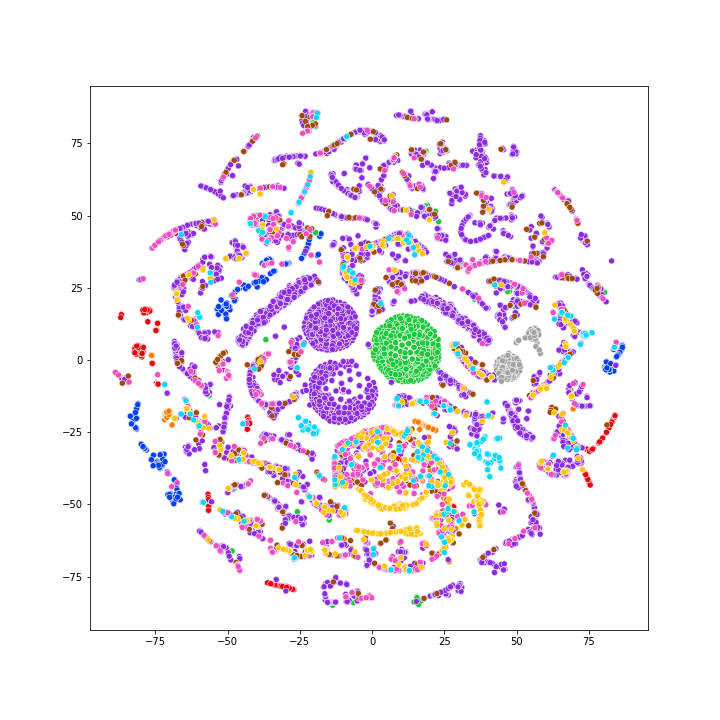}
%\vskip -0.5cm
\caption{EMBER 10-Class}
\label{fig:ember_class_10_tSNE}
\end{subfigure}%
\vskip -0.1cm
\caption{Feature space visualization using t-SNE in Class-IL scenario.}
\label{fig:featspacevisualize}
\vskip -0.3cm
\end{figure}

Continual learning (CL) for malware classification can provide a number of benefits including: i) alleviating the need to store some or all of the previous data, ii) relaxing the requirement for access to previous data, iii) keeping training data private while sharing a model that can be updated, and iv) reducing computational cost. 
%by replaying partial stored data while using the stored data. 
From our experiments, however, we can see that none of the CL techniques could contribute significantly in the Domain-IL setting, and only iCaRL provided reasonable performance in Class-IL setting, though even in that setting the gap between {\em Joint} and iCaRL was substantial -- upwards of 24\% on EMBER. For Task-IL, we observed reasonable performance by several methods on Drebin and on EMBER. Task-IL for malware, however, is of least significance, as adding new tasks is likely to occur less frequently than adding new classes or observing domain shift.
Given these limitations, we now examine probable causes to help spur future research in CL considering these problem settings and datasets.

\paragraph{Dataset Complexity.} Most papers on CL techniques use the MNIST dataset for their experiments, but MNIST is relatively simple in its data distribution and feature space. Figure~\ref{fig:featspacevisualize} shows a t-SNE projection~\citep{van2008visualizing} of the feature space to demonstrate the complexity of MNIST, Drebin, and EMBER dataset with both two and 10 classes for each. In MNIST, we can see a clear separation for the two classes and relatively clear separations for 10 classes. In Drebin, we see a somewhat sparser, yet complex space, particularly for 10 classes where some classes become difficult to separate. Meanwhile, the EMBER feature space is extremely complex with significant overlap between classes. Additionally, we see class imbalance in the EMBER dataset, which reflects the realistic sampling of real-world data. MNIST has equal numbers of samples in both Class-IL and Task-IL.

Another interesting distinction is the semantically-rich feature space of the malware classification datasets. Unlike the image domain, malware classifiers typically use tabular features derived from parsing and analysis of the binary program (e.g., headers, byte sequences, APIs). There are strong semantic constraints on the values that these features can take on and their relationships, which affects the shape of the feature space that feasible samples occupy.  It is possible that these constraints and the inherent complexity of the classification task render generative, distillation, and regularization-based CL methods ineffective. Taken together, our  results show that it is important to evaluate CL techniques on more complex, realistic datasets to understand how their effectiveness may generalize and apply in practical settings.

%We can observe the collusion of the feature space is much more intense in EMBER dataset in incremental classes of 2 and 10. Hence, we emphasize that feature space complexity should be an important consideration for future study in this problem setting.

%The feature space of Drebin is relatively simpler than EMBER, but more complex than MNIST, while the feature space of EMBER is significantly complex than MNIST. 

%The assumption made in the study of Class-IL scenario using {\em split} MNIST protocol where each task represents an equal number of incremental classes is very strong. In a realistic Class-IL scenario, we would have a well performed base model with our current $N$-number classes, and then we would increment any number of classes in the base model.  Prior work weighed on the fact that the performance of CL techniques degrades with incremental task~\citep{van2020brain}. We argue that data complexity is also an important property that lead to a performance drop task by task.

\paragraph{Real Domain Shifts.} Prior work has studied Domain-IL only in a simulated experimental setting to mimic data distribution shift using the {\em permuted MNIST} protocol, in which pixels of the MNIST images are permuted around the image in a consistent way across the dataset to create the next task. This protocol clearly does not represent a realistic data distribution shift, and certainly not the strongly constrained feature space found in malware classification tasks. As such, the performance offered by CL techniques in this simulated setting is not a realistic measurement that would be likely to generalize to other data. In contrast, our experimental setting in Domain-IL represents a sample of real-world data that shifts over time. 
%We can see that none of the CL techniques can offer notable improvements over {\em None} benchmark in this setting which indicate that studied CL techniques lack the capacity to carry the knowledge of the older task to reduce catastrophic forgetting. When the feature space becomes more colluded with more and more distributional shifts in incremental domains, it becomes harder for the regularized CL techniques to retain the weights that are important for the previously learned domain. 
In Appendix~\ref{domain_feature_space}, we show a t-SNE visualization of MNIST and EMBER data distribution shift in Domain-IL. In summary, even though permuted MNIST is challenging for people to visualize, it appears to actually create reasonably solvable classification tasks for DL models. The natural distribution shift found in EMBER, on the other hand, creates very difficult tasks for the classifier.

\paragraph{Partial Joint Replay (PJR).}
Simple partial joint replay offers significant improvements over the {\em None} benchmark while also reducing computational effort. It would be interesting for future work to evaluate the effectiveness of combining partial joint replay and CL techniques, and our results here offer a pragmatic benchmark for future efforts. 
In some ways, the widely studied iCaRL CL technique is analogous to PJR in that iCaRL replays the samples stored in the memory buffer in the training phase, coupling those old samples with the new ones. %Given a memory budget (i.e., total number of exemplars that can be stored), $\mathcal{K}$, iCaRL stores and uses $\mathcal{M} = \mathcal{K}/\mathcal{C}$, where $\mathcal{C}$ is the number of classes observed so far. The pitfall of this technique is that the more classes that are learned, the fewer number of exemplars in each class remain in the memory buffer due to the fixed memory budget. Unless the number of new classes can be determined in advance, to provide a suitable memory budget,
%As the number of new classes is not deterministic to choose a greater $\mathcal{K}$ to ensure that enough number of exemplars of the learned classes remain in the buffer, 
As originally designed, however, iCaRL has a fixed memory budget that can limit its effectiveness with increasing classes. We can validate this by observing the performance of iCaRL for Drebin and EMBER datasets shown in Table~\ref{exp_summary} and Figures~\ref{fig:drebin_exp_class} and~\ref{fig:ember_exp_class}. Drebin has 18 classes and the performance gap between iCaRL and {\em Joint} is around 3\%. On the other hand, EMBER has 100 classes and iCaRL's performance gap grows to almost 25\%. 

\section{Conclusion}

%The current approaches to train a machine learning (ML) and deep learning (DL) based malware classification assume that the training and the test distribution are static, however, in practice the training distribution may change, new families may evolve, new samples are identified. Retraining with all the previous data and the new data is the current approach to retain the upper bound performance of the system, otherwise the model will suffer from {\em catastrophic forgetting} -- a phenomena in which the model forgets all the previously learned tasks. Retraining is, however, expensive with the periodic increase of the size of data and previous data may become unavailable. In addition, biological neurons learn in a sequential manner and do not forget the older tasks. {\em Continual learning} attempts to overcome catastrophic forgetting in a model while training the model sequentially.
In this work, we investigate 11 continual learning techniques in three scenarios using two large, real-world malware datasets. Unfortunately, our findings demonstrate that in almost all cases those CL techniques are ineffective at preventing catastrophic forgetting and maintaining classification performance. Of the techniques evaluated, only iCaRL~\citep{icarl} performed near {\em Joint} replay baselines. Meanwhile, we also found that a simple partial joint replay strategy of training on a small fraction of the historical data is enough to achieve reasonable performance with lower cost.

Taken as a whole, our results underscore the need for additional study of CL in complex, non-stationary problem settings. We hypothesize that the strong semantic constraints of the features used for malware classification tasks, along with the complex data distributions induced by natural drift in this space, lie at the heart of the performance differences between existing CL literature and the results in our paper. In particular, these unique aspects of the malware classification domain may severely limit the applicability of generative, distillation, and regularization-based methods due to their inability to sufficiently capture the inherent complexity. At the same time, these results also hint at possible avenues of future work. For example, the relative success of partial joint replay and iCaRL demonstrate the importance of sample selection during replay and the possibility of developing more effective strategies that ensure optimal coverage over feasible regions of the feature space. Most importantly, however, we hope this work spurs further exploration of CL in the cybersecurity domain and a broader conversation about real-world application of CL techniques.

\balance

%\if 0
\paragraph{Acknowledgements.}
We thank the anonymous reviewers for their helpful feedback. This material is based upon work supported in part by the National Science Foundation under Grant No.~1816851.
%\fi

\bibliography{main}
\bibliographystyle{collas2022_conference}

\appendix

\section{Dataset details}

\paragraph{Drebin.}
\label{drebindatastat}
We can see the details of Drebin dataset used in this study from Table~\ref{drebin_data}. The total number of samples of the top 18 classes are 4525. Unfortunately, we could not find the samples belonging to {\em LinuxLotoor} and {\em GoldDream} in the dataset.

\begin{table}[ht]
\centering

\caption{\textbf{Drebin.} 4525 malware samples from top 18 malware families.}
\begin{tabular}{l|l|c||l|l|c} 
\textbf{Label} & \textbf{Family} & \textbf{\#} & \textbf{Label} & \textbf{Family} & \textbf{\#}\\ 
\hline
%\rule{0pt}{2ex}
0   &   FakeInstaller   &925    &9&     Geinimi&   92 \\
1   &   DroidKungFu     &667    &10&    Adrd&   91\\
2   &   Plankton        &625    &11&    DroidDream& 81\\
3   &   Opfake          &613    &12&    MobileTx&   69\\
4   &   GingerMaster       &339    &13&    FakeRun&    61\\
5   &   BaseBridge      &330    &14&    SendPay&    59\\
6   &   Iconosys        &152    &15&    Gappusin&   58\\
7   &   Kmin            &147    &16&    Imlog&  43\\
8   &   FakeDoc         &132    &17&    SMSreg& 41\\
\hline \hline
%\rule{1pt}{3ex}
\multicolumn{6}{c}{\textbf{Total 4,525}}\\
\hline
%\rule{0pt}{3ex}
\end{tabular}
\label{drebin_data}
%\vskip -0.4cm   
\end{table}

%Preparing Drebin malware data...
%Family FakeInstaller has 925 samples
%Family DroidKungFu has 667 samples
%Family Plankton has 625 samples
%Family Opfake has 613 samples
%Family GinMaster has 339 samples
%Family BaseBridge has 330 samples
%Family Iconosys has 152 samples
%Family Kmin has 147 samples
%Family FakeDoc has 132 samples
%Family Geinimi has 92 samples
%Family Adrd has 91 samples
%Family DroidDream has 81 samples
%Family MobileTx has 69 samples
%Family FakeRun has 61 samples
%Family SendPay has 59 samples
%Family Gappusin has 58 samples
%Family Imlog has 43 samples
%Family SMSreg has 41 samples
%train_shas: 3627, test_shas: 898
%Number of Training features before Variance Thresholding 8803
%Before feature selection X shape: (3627, 8803)
%After feature selection X_select shape: (3627, 2492)
%X_train_final : (3627, 2492), y_train_final: (3627,)
%X_test_final: (898, 2492), y_test_final: (898,)
%total data 4525

\paragraph{EMBER.}
\label{emberdatastat}

\begin{figure}[!ht]
%\vskip -1.0cm
\centering
\includegraphics[scale=0.40]{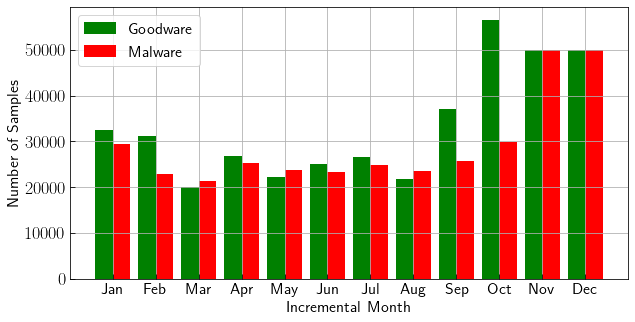}
%\vskip 0.2cm    
\caption{\textit{\bf EMBER Data}: Goodware and malware data statics of 12 months of 2018.}
\label{fig:ember_data_stat}
%\vspace{-0.4cm}
\end{figure}

Figure~\ref{fig:ember_data_stat} shows the number of goodware and malware samples of each month of 2018.

\section{Details of the Adapted Continual Learning Techniques}
\label{adaptedtechniquesdetails}

\paragraph{Elastic Weight Consolidation (EWC) and EWC Online.} \citet{ewc} proposed EWC to overcome catastrophic forgetting in a neural network. EWC is inspired by human brain, in which the plasticity of the synapses of the previously learned tasks are reduced to facilitate continual learning. As mentioned earlier, excessive plasticity of the weights of the previous tasks is the major cause of the catastrophic forgetting. If the plasticity of the weights is loosely connected to the previous task, then the network becomes less prone to catastrophic forgetting. Leveraging this idea, EWC quantifies the importance of the weights in terms of their impact on the previous tasks' performance, and selectively reduces the plasticity of the most important such weights.

A Bayesian approach is used to measure the importance of the parameters of a task in EWC. Given two tasks $T_1$ and $T_2$, EWC tries to converge to a point where both of these tasks have low error using the following loss function:

\begin{equation}\label{equationewc}
    \mathcal{L}(\theta) = \mathcal{L}_{T_2}(\theta) + \sum_{i}\dfrac{\lambda}{2}F_{i}\left(\theta_{i} - \theta_{T_{1},i}^*\right)^2 
\end{equation}

\noindent Here, $\mathcal{L}_{T_{2}}(\theta)$ is the loss of task $T_2$ only, $F_{i}(\theta_{i} - \theta_{T_1,i}^*)^2$ approximates a Gaussian distribution with mean given by the parameters $\theta_{T_1}^*$ which is with respect to task $T_1$ and a diagonal of the Fisher information matrix $F$. $i$ is an index into the weight vector. $\lambda$ controls this distribution in such as way that the weights do not move too far away from the low error region of task $T_1$. Similarly, when a new task $T_3$ is observed, the loss function is updated in such a way that forces the parameters $\theta$ to be close to $\theta_{T_1,T_2}^*$, where $\theta_{T_1,T_2}^*$ is the parameters learned for the previous tasks $T_1$ and $T_2$.

%\paragraph{EWC Online}
The major drawback of the proposed EWC is scalability. The quadratic term of the regularization loss grows linearly with the increase of task which hinders to enable EWC be truly applicable in a practical continual learning scenario where the tasks keep growing. EWC Online attempts solve this problem by strict Bayesian treatment which results in a single quadratic penalty term considering the important parameters of the current task and a running sum of the Fisher Information matrices of the previous task's parameters~\citep{onlineewc}. EWC Online modifies the second part of Equation~\ref{equationewc} where $\tilde{F}_{i}$ is the running sum of the previous task. The modified loss function becomes as follows:

\begin{equation}\label{equationewconline}
    \mathcal{L}(\theta) = \mathcal{L}_{T_2}(\theta) + \sum_{i}\lambda \tilde{F}_{i}\left(\theta_{i} - \theta_{T_{1},i}^*\right)^2 
\end{equation}

\paragraph{Synaptic Intelligence (SI)} SI is a variant of EWC, in which changes to weights that are important to the previous tasks are penalized when training on new tasks~\citep{si}. Fundamentally, regularizer loss is added to the loss of the current task to get the total loss, $\mathcal{L}_{total}$. To compute this regularization loss, we first compute the importance of the weights $I_{t}^w$ after every task $t$, with respect to the change in total loss $\mathcal{L}_{total}$. To get the estimated importance of the weights $I_{t}^{N-1}$ for $N-1$ tasks, all the $I_{t}^w$ are summed up together after a normalization step.
%by the square of the total change of weights during training plus a small dampening terms $\gamma$ which is set to $0.1$ so that the normalized contributions can be bound when weights' change goes to zero. 
Finally, the regularization loss $\mathcal{L}_r(\theta)$ for tasks $N > 1$ is computed in Equation~\ref{SIformula}. Refer to the paper~\citep{si} for more details.

\begin{equation}\label{SIformula}
    \mathcal{L}_r(\theta) = \sum_{t=1}^{K} I_{t}^{N-1} \left(\theta_t - \hat\theta_{t}^{(N-1)}\right)^2
\end{equation}

\paragraph{Experience Replay (ER)} 
ER technique jointly trains the network utilizing both the examples (i.e., data) from the current task and examples stored in the very small episodic
memory. ER is based on a distributed actor-critic training in which the single learner is fed both new and replayed experiences (i.e., tasks)~\citep{er}. To adjust the off-policy distribution shifts during training, ER adapts V-Trace off-policy learning algorithm~\citep{impala}. Three losses -- $L_{policy-gradient}, L_{value},$ and $L_{entropy}$ are common to both new and replayed experiences and another two losses -- $L_{policy-cloning}$ and $L_{value-cloning}$ are unique to only the replayed experiences. Refer to the paper~\citep{er} for more details.

\paragraph{Learning without Forgetting (LwF)}
LwF tries to reduce the catastrophic forgetting of an older task $t_{n-1}$ by learning the parameters of new task $t_n$, $\theta_n$, with the help of the shared parameters $\theta_s$ and the parameters of the older tasks $\theta_0$. The objective is to optimize $\theta_n$ and $\theta_s$ such that the prediction of $t_n$ using $\theta_s$ and $\theta_0$ does not drift significantly~\citep{lwf}. The objective function of LwF algorithm is given below:

\begin{equation}
    \begin{split}
        \theta_{s}^*, \theta_{0}^*, \theta_{n}^* \longleftarrow \operatorname{argmin}_{\hat \theta_{s}, \hat \theta_{0}, \hat \theta_{n}} (\mathcal{L}_{new}(\hat y_{n}, y_{n})  + \\ \lambda_0\mathcal{L}_{old}(\hat y_{0}, y_{0}) + \mathcal{R}(\theta_s,\theta_0, \theta_n))
    \end{split}
\end{equation}

\noindent $\hat y_n$ is the prediction of the test samples of the new task $t_n$ using current shared parameters $\hat \theta_s$ and current task's parameters $\hat \theta_n$. A multinomial logistic loss function is used to compute $\mathcal{L}_{new}$. Thus, $\mathcal{L}_{new}(\hat y_{n}, y_{n})$ minimizes the difference between predicted $\hat y_n$ and actual $y_n$. $y_0$ is the prediction of the test sample of the new task $t_n$ using the shared parameters $\theta_s$ and previous task's parameters $\theta_0$ (i.e., the model before being trained with the samples of the new task). $\hat y_0$, on the other hand, is the prediction of the test samples of the new task $t_n$ using current shared parameters $\hat \theta_s$ and previous task's parameters $\hat \theta_0$ (i.e., the model as trained with the samples of the new task). Distillation loss~\citep{hinton2015distilling} is used to compute the $\mathcal{L}_{new}$ so that the output of one network can be approximated using the outputs of another. Thus $\mathcal{L}_{old}(\hat y_{0}, y_{0})$ minimizes the difference between $\hat y_0$ and $y_0$. $\lambda_0$ works as a balancing factor between the new task $t_n$ and the previous task $t_{n-1}$. In the algorithm, $\mathcal{R}(\theta_s,\theta_0, \theta_n)$ works as a regularizer to avoid overfitting in the model.

\begin{figure}[!t]
\vskip -1.0cm
\begin{minipage}[c]{1.0\linewidth}
\centering
\includegraphics[scale=0.45]{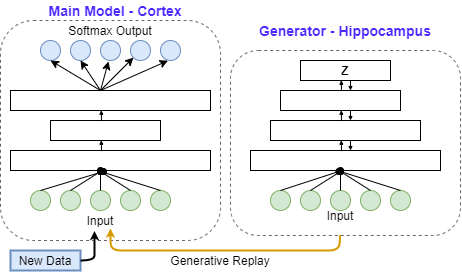}
\caption{A schematic representation of Generative Replay.}
\label{fig:generativereplay}
\end{minipage}
\end{figure}

\paragraph{Generative Replay (GR) and GR with Distillation}
Some of the earlier literature suggests that the cerebral cortex, which works as the main model, works better when paired with a generative model than a replay buffer~\citep{killgore2008sleep, ramirez2013creating,chen2018lifelong}. In 2017, Shin et al.~\citep{gr} proposed Deep Generative Replay (DGR), in which the replayed examples of the previous tasks are generated using a Generative Adversarial Network (GAN)~\citep{goodfellow2014generative}, meaning that real data from the past tasks do not need to be stored. We can see a schematic representation of GR in Figure~\ref{fig:generativereplay}, where the generator works as the hippocampus and generates the representative samples of the stored data and replayed during the training of the model with new data. The GAN generator creates data to represent the model's knowledge of the previous tasks.

Given a series of tasks $t_1, t_2,...,t_n$, an expert model $\mathcal{S}$ -- which contains a generator model $\mathcal{G}$ and a solver model $\mathcal{M}$ (called the main model in some works~\citep{rtf, bir}) -- holds the knowledge of the previous tasks and thus prevents the system from catastrophic forgetting. In this work, we adapt the DGR framework for our malware classification problem.

In DGR, $\mathcal{S}$ is learned and maintained in a continual learning fashion. $\mathcal{S}$ for the series of previous $n$ tasks can be represented as $\mathcal{S}_n = (\mathcal{G}_n, \mathcal{M}_n)$. Given a new task $t_{n+1}$ and the new training data $D_{n+1}$, the objective of $\mathcal{S}$ is to learn $\mathcal{S}_{n+1} = (\mathcal{G}_{n+1}, \mathcal{M}_{n+1})$. There are two steps involved in the learning process of $\mathcal{S}_{n+1}$ considering $D_{n+1} = (x,y)$ where $(x,y)$ represents {\em (data, label)}:

\begin{enumerate}
    \item At first, the {\em scholar} model $\mathcal{S}_{n+1}$ is updated with the input $x$ of new task $t_{n+1}$ and replayed with the generated data, $\hat x$, from previous scholar model $\mathcal{G}_n$. Real data $x$ and replayed data $\hat x$ are mixed together at a ratio based on the importance of the new task $t_{n+1}$ compared to the older task $t_n$. This is referred to as intrinsic replay or pseudo-rehearsal~\citep{robins1995catastrophic}.
    \item Then the main model $\mathcal{M}_{n+1})$ is trained with the real and replayed data with the following loss function:
    \begin{equation}
        \begin{split}
            \mathcal{L}_{train} (\theta_{n+1}) = r\mathop{\mathbb{E}}(x,y)\sim\mathcal{D}_{n+1}[\mathcal{L}(\mathcal{M}(x;\theta_{n+1}),y)] +\\
            (1 - r) \mathop{\mathbb{E}}(x,y)\sim\mathcal{D}_{n}[\mathcal{L}(\mathcal{M}(x;\theta_{n+1}),y)]
        \end{split}
    \end{equation}
    
    Here, $\theta_n$ represents the parameters of the main model $\mathcal{M}_n)$ and $r$ represents the ratio of the mixture of the real and replayed data.
\end{enumerate}

The DGR framework is designed in such a way that choice of the generative model is not limited to a GAN and can instead be a variational autoencoder (VAE)~\citep{kingma2013auto} or any other such type.

%\subsection{Generative Replay (GR)}
For the experiments of GR, we need two models -- the main model $\mathcal{M}$ and a generative model $\mathcal{G}$. $\mathcal{G}$ is responsible for generating representative samples of the previous tasks to be replayed in the current task. We use the base model architecture for both $\mathcal{M}$ and $\mathcal{G}$. The loss function of $\mathcal{M}$ consists of two parts -- one for the data of the current task and another for the replayed samples. The cumulative loss of these two parts are weighted in terms of the number of tasks the model has observed so far.

For $\mathcal{G}$ in our experiments, we use a symmetric VAE~\citep{kingma2013auto}, where there is an encoder that maps the input data distribution to a latent distribution and a decoder that reconstructs the inputs from the latent distribution. For both the encoder and the decoder, the base model architecture is used. In all of our experiments, we used a stochastic latent variable layer with 100 Gaussian units parameterized by the mean and the standard deviation of the output of the encoder given input $x$.

%\paragraph{Replayed Samples.}
The data to be replayed are sampled from the generative model, and then the selected samples are fed to the main model and then labeled based on the predicted class of the model. The samples to be replayed during task $T$ are generated by the version of the main model and the generator after training on task $T-1$. Hence, we need to store a copy of both $\mathcal{M}$ and $\mathcal{G}$ after each task. 

GR with distillation is a variant of GR where the generated samples are replayed with the output probabilities (i.e., soft targets) instead of the actual labels. Previous work show that GR with distillation often works better than GR~\citep{van2019three, rtf, bir}.

\paragraph{Replay through Feedback (RtF)}
GR has two models -- a main model and a generative model. RtF proposes to merge the generator model into the main model~\citep{rtf}. This is inspired by the fact that replay in the brain is originated in the hippocampus and then it propagates to the cortex, and in our brain's processing hierarchy, the hippocampus sits on top of the cortex. The merged model will work as a brain, where the first $n$ layers will work as the visual cortex and the last $m-n$ layers will work as the hippocampus. Technically, an additional \texttt{softmax} classification layer is added on top of the encoder of our generator VAE model. This technique requires only a single model to be trained, and the loss of the current tasks has two terms -- cross entropy loss and generative loss.

\paragraph{Brain-Inspired Replay (BI-R)}
Recently proposed by~\citet{bir}, BI-R seeks to improve upon RtF with another three add-on components -- Conditional Replay (CR), Gating based on Internal Context (Gating), and Internal Replay (IR). For CR, BI-R proposes to replace the standard normal prior over the VAE's latent variables by a Gaussian mixture with a separate mode for each class so that class-specific samples can be generated. This is due to the fact that a vanilla VAE is limited to generate class-specific samples, but humans do have control over which memories to recall. Conditional replay (CR) is intended to provide the network a human-like capacity to generate samples of the class the network needs most.
Context-dependent gating was originally proposed by~\citet{xdg}. The idea is to reduce interference between different tasks by gating different and randomly selected network nodes for each task. However, this technique requires the task identity to be known for all the tasks, which is not realistic in the Class-IL scenario. BI-R proposes to use this gating technique in the decoder of the VAE with a conditional of the internal context. The task or class to be generated and reconstructed is the conditioned internal context. To note, not all of the nodes of the decoder network are gated.
Mental images are not propagated all the way to the retina, and thus the brain does not replay memories to the input level~\citep{bendor2012biasing}. This insight is corroborated by evidence from neuroscience that our brain's early visual cortex does not change significantly from childhood to adulthood~\citep{smirnakis2005lack}. To accommodate these observations into continual learning, BI-R proposes to replay internally or at a hidden layer, instead of to the input level. From the machine learning perspective, the first $n$ layers will need a limited amount of change, since there is no replay in them.

\paragraph{Incremental Classifier
and Representation Learning (iCaRL)}
iCaRL~\citep{icarl} is one of the earliest replay-based methods specifically designed for Class-IL scenario. Given a fixed buffer size (i.e., allocated memory), iCaRL stores samples of the earlier learned classes which are closest to the feature mean of those classes obtained from the feature maps of the network. iCaRL minimizes two loss functions -- i) one is the categorical cross entropy loss of new classes, and ii) distillation loss obtained from the predictions of the current model's and the previous model's targets.

\paragraph{Averaged Gradient Episodic Memory (A-GEM)} A-GEM~\citep{agem} is an improved version of GEM\citep{gem}. GEM attempts to reduce catastrophic forgetting by constraining the updates of the new task not to interfere with the previous tasks. GEM utilizes the first order Taylor series approximation to estimate the direction of the gradient on the possible areas laid out by the gradients of the previously learned tasks. A-GEM relaxes the constraint to project the gradient into only one direction estimated from the randomly selected samples stored in a replay buffer. The replay buffer contains samples of the previously learned tasks.

\section{Feature space progression in Domain-IL Scenario}
\label{domain_feature_space}

\begin{figure}[!t]
%\vskip -.50cm
\centering
\begin{subfigure}[c]{0.32\linewidth}
\centering
\includegraphics[scale=0.20]{figures/mnist_tSNE_new_01.png}
\vskip -0.5cm
\caption{\textit{\bf Task 1}: Original MNIST (no permutation).}
\label{fig:mnist_perm0_tSNE}
\end{subfigure}%
\hfill
\begin{subfigure}[c]{0.32\linewidth}
\includegraphics[scale=0.20]{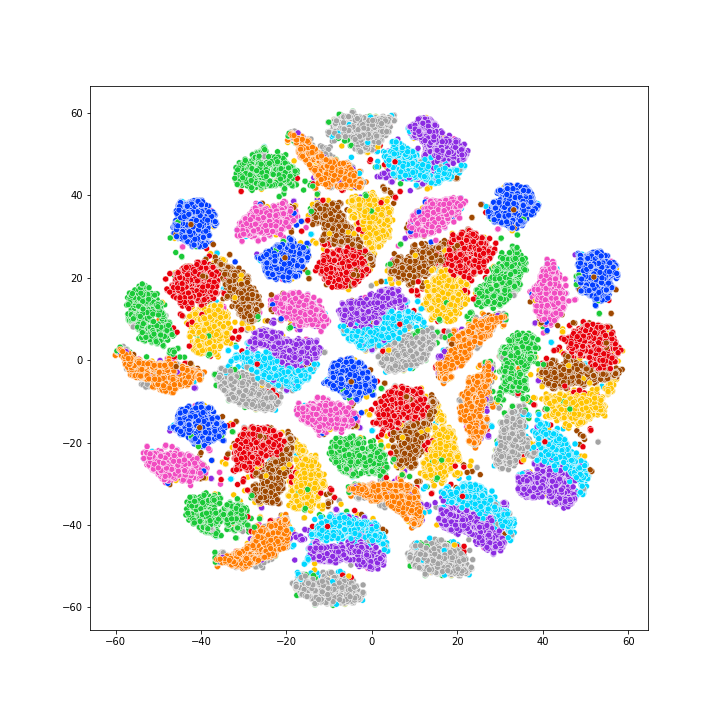}
\vskip -0.5cm
\caption{\textit{\bf Task 6}: Cumulative MNIST data from Task 1 to Task 6 using {\em permuted} MNIST protocol.}
\label{fig:mnist_perm6_tSNE}
\end{subfigure}
\hfill
\begin{subfigure}[c]{0.32\linewidth}
\centering
\includegraphics[scale=0.20]{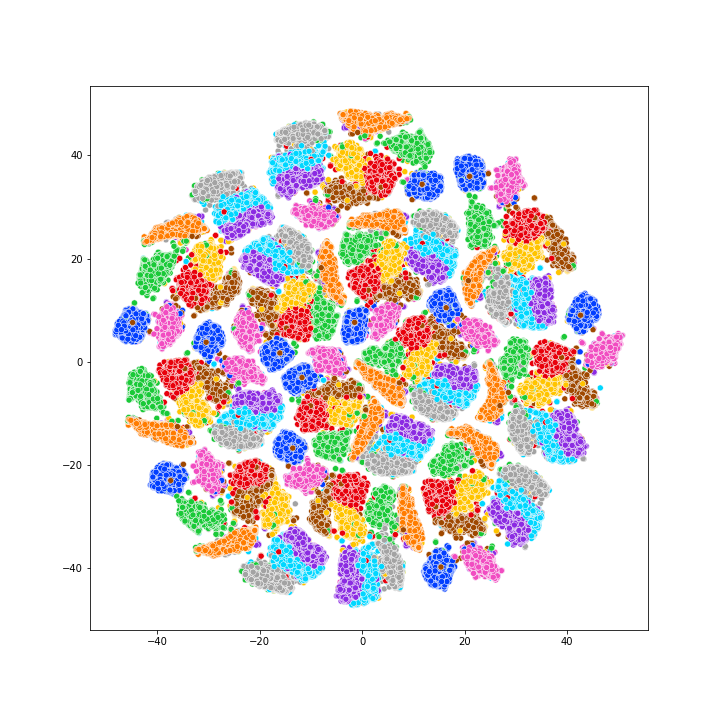}
\vskip -0.5cm
\caption{\textit{\bf Task 12}: Cumulative MNIST data from Task 1 to Task 12 using {\em permuted} MNIST protocol.}
\label{fig:mnist_perm3_tSNE}
\end{subfigure}%
\vfill
\begin{subfigure}[c]{0.32\linewidth}
\centering
\includegraphics[scale=0.20]{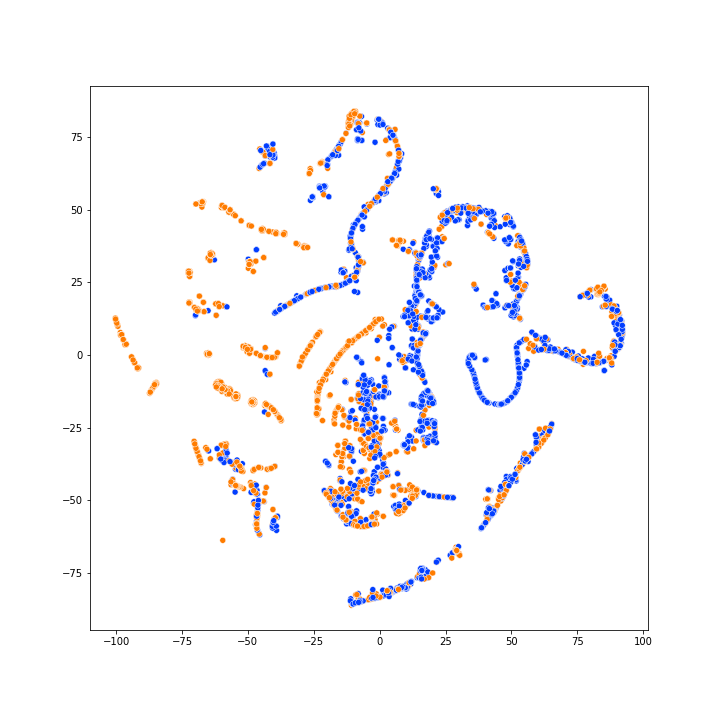}
\vskip -0.5cm
\caption{\textit{\bf Task 1}: EMBER data of January}
\label{fig:ember_jan_tSNE}
\end{subfigure}%
\hfill
\begin{subfigure}[c]{0.32\linewidth}
\includegraphics[scale=0.20]{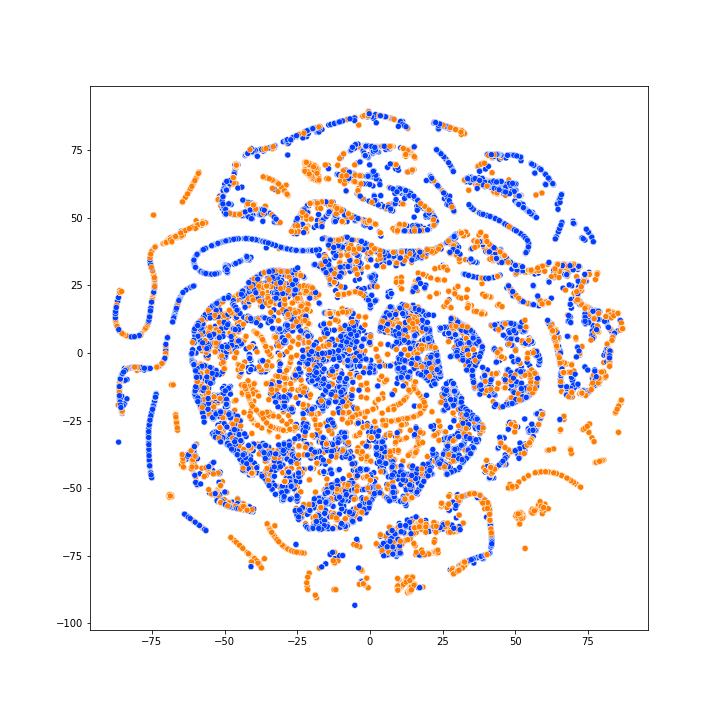}
\vskip -0.5cm
\caption{\textit{\bf Task 6}: Accumulated EMBER data from January to June.}
\label{fig:ember_jan_june_tSNE}
\end{subfigure}
\hfill
\begin{subfigure}[c]{0.32\linewidth}
\centering
\includegraphics[scale=0.20]{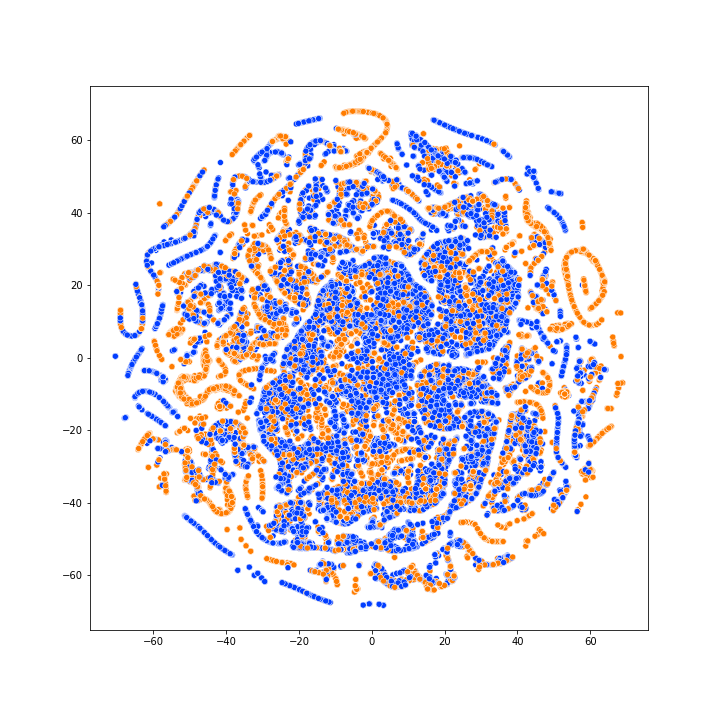}
\vskip -0.5cm
\caption{\textit{\bf Task 12}: Accumulated EMBER data from January to December.}
\label{fig:ember_jan_dec_tSNE}
\end{subfigure}%
\caption{MNIST and EMBER data distribution shift in Domain-IL scenario using t-SNE plot.}
\label{domain_data_distribution}
\end{figure}
%\section{Appendix}
%You may include other additional sections here.

\end{document}